\documentclass[apj, letterpaper]{emulateapj}  

\usepackage{apjfonts}  

\usepackage{amsmath}

\numberwithin{equation}{section}

\newcommand{\m}{\rm \,m}

\newcommand{\J}{\rm \,J}
\newcommand{\kg}{\rm \,kg}

\newcommand{\W}{\rm \, W}

\newcommand{\km}{\rm \, km}

\begin{document}

\slugcomment{\bf}
\slugcomment{Submitted to the Astrophysical Journal}

\title{Atmospheric circulation of hot Jupiters: Coupled 
radiative-dynamical general circulation model simulations of
HD 189733b and HD 209458b }

\shorttitle{Atmospheric Circulation of Hot Jupiters}

\shortauthors{Showman et al.}

\author{Adam P.\ Showman\altaffilmark{1}, 
Jonathan J.\ Fortney\altaffilmark{2}, Yuan Lian\altaffilmark{1},
Mark S. Marley\altaffilmark{3}, Richard S. Freedman\altaffilmark{3,4},
Heather A. Knutson\altaffilmark{5},
and David Charbonneau\altaffilmark{5}}

\altaffiltext{1}{Department of Planetary Sciences and Lunar and Planetary
Laboratory, The University of Arizona, 1629 University Blvd., Tucson, AZ 85721 USA; showman@lpl.arizona.edu}
\altaffiltext{2}{Department of Astronomy \& Astrophysics, University of California, Santa Cruz, CA 95064}
\altaffiltext{3}{NASA Ames Research Center 245-3, Moffett Field, CA 94035}
\altaffiltext{4}{SETI Institute, 515 North Whisman Road, Mountain View, CA 94043}
\altaffiltext{5}{Harvard-Smithsonian Center for Astrophysics, 
60 Garden Street, Cambridge, MA 02138}

\begin{abstract}
\label{abstract}

We present global, three-dimensional numerical simulations of
HD 189733b and HD 209458b that couple the atmospheric dynamics to a realistic
representation of non-gray cloud-free radiative transfer.  The model,
which we call the Substellar and Planetary Atmospheric Radiation and
Circulation (SPARC) model,
adopts the MITgcm for the dynamics and uses the 
radiative model of McKay, Marley, Fortney, and collaborators
for the radiation.   Like earlier work with simplified forcing, 
our simulations develop a broad eastward equatorial 
jet, mean westward flow at higher latitudes, and substantial 
flow over the poles at low pressure.  For HD 189733b, our simulations 
without TiO and VO opacity can explain the broad features of the observed 8 and
24-$\mu$m light curves, including the modest day-night flux variation
and the fact that the planet/star flux ratio peaks before the secondary
eclipse.  Our simulations also provide reasonable matches to the 
{\it Spitzer} secondary-eclipse depths at 4.5, 5.8, 8, 16, and $24\,\mu$m 
and the groundbased upper limit at $2.2\,\mu$m.  However, we substantially 
underpredict the  $3.6\,\mu$m secondary-eclipse depth, suggesting that
our simulations are too cold in the 0.1--1 bar region.  Predicted 
temporal variability in secondary-eclipse depths is $\sim1$\% at
{\it Spitzer} bandpasses, consistent with recent observational upper limits
at $8\,\mu$m.  We also show
that nonsynchronous rotation can significantly alter the jet structure.
For HD 209458b, we include TiO and VO opacity; these simulations develop
a hot ($>2000\,$K) dayside stratosphere whose horizontal dimensions 
are small at depth but widen with altitude.  Despite this stratosphere, 
we do not reproduce current {\it Spitzer} photometry of this planet.   
Light curves in {\it Spitzer} 
bandpasses show modest phase variation and satisfy the observational
upper limit on day-night phase variation at $8\,\mu$m.

\end{abstract}



\section{Introduction}
\label{Introduction}

Blasted by starlight $10^3$--$10^5$ times stronger than 
that received by Jupiter and experiencing modest rotation
rates due to their presumed tidal locking \citep{guillot-etal-1996},
hot Jupiters occupy a fascinating meteorological regime that does
not exist in our Solar System \citep[for an extensive review see][]
{showman-etal-2008b}. 
Despite the wide range of transiting exoplanets that have been 
discovered, HD 189733b and HD 209458b remain the best characterized
hot Jupiters and represent important test cases for our understanding of these 
objects generally.  A variety of observations now exist that constrain
the 3D temperature structure, composition, hazes, and albedo 
for these two planets.  There is now hope that, by comparing these
observations with detailed models, insights into this novel atmospheric
regime can be achieved.

The strongest observational evidence for atmospheric motions comes from
infrared light curves obtained with the {\it Spitzer Space Telescope}.
Continuous light curves of HD 189733b over half an orbit have now 
been obtained at both 8 and $24\,\mu$m, constraining this planet's 
day-night heat transport \citep{knutson-etal-2007b, knutson-etal-2009a}. 
The inferred dayside and nightside brightness temperatures 
are $\sim1250\,$K and $\sim1000\,$K, respectively.\footnote{At $8\,\mu$m,
the dayside and nightside brightness temperatures are
$1258\pm11\,$K and $1011\pm51\,$K.  At $24\,\mu$m, the dayside and 
nightside brightness temperatures are $1220\pm47\,$K and 
$984\pm48\,$K.}
Because the nightside temperatures
would be extremely cold ($\sim200\,$K) in the absence of winds, 
these observations imply the existence
of a vigorous atmospheric circulation that efficiently transports heat 
from dayside to nightside. Further evidence for winds is the fact that 
the peak flux in both light curves occurs {\it before}
secondary eclipse, implying that the hottest region lies 20--$30^{\circ}$
of longitude east of the substellar point \citep{knutson-etal-2007b,
knutson-etal-2009a}.
For HD 209458b, current data contain insufficient temporal sampling
to determine whether similar offsets exist but nevertheless demonstrate
that the 8-$\mu$m day-night brightness temperature difference is also 
modest \citep{cowan-etal-2007}.

For both planets we now also
have dayside photometry at all Spitzer broadband channels 
(centered at 3.6, 4.5, 5.8, 8.0, 16, and $24\,\mu$m), 
placing constraints on dayside composition and vertical temperature 
profile.  These data were obtained by differencing the photometry 
of star$+$planet taken just before/after secondary eclipse from 
photometry taken during secondary eclipse, when only the star is visible.  
When compared with one-dimensional (1D) atmosphere models, 
these data suggest that, near the photosphere pressures, the dayside 
temperature of HD 189733b decreases with altitude \citep{charbonneau-etal-2008,
barman-2008, knutson-etal-2009a}, whereas HD 209458b instead 
contains a thermal inversion (a hot stratosphere) \citep{knutson-etal-2008a, 
burrows-etal-2007b}.

Following pioneering work by \citet{hubeny-etal-2003},
\citet{fortney-etal-2008} and \citet{burrows-etal-2008} suggested 
that hot Jupiters subdivide into two classes depending on whether or 
not their atmospheres contain highly absorbing substances such as
gaseous TiO and VO. For a Sun-like primary, solar-composition planetary
atmospheres inward of 0.04--$0.05\,$AU are hot enough to contain TiO and VO; 
because of the extreme visible-wavelength opacity of these compounds, 
such planets 
absorb starlight at low pressure ($\sim$mbar) and naturally exhibit dayside 
stratospheres.  For planets outward of $\sim0.05\,$AU, temperatures 
drop sufficiently for TiO and VO to condense in the deep atmospheres; these 
planets absorb starlight deeper in the atmosphere, lack stratospheres, 
and show spectral bands in absorption.    
Based on simple comparisons of radiative and advective 
timescales, \citet{fortney-etal-2008} further suggested 
that the former category
(dubbed ``pM'' class planets) would exhibit large day-night temperature 
differences whereas the latter category (``pL'' class) would exhibit 
more modest temperature contrasts.  Their calculations suggest that 
HD 189733b is a pL-class planet while HD 209458b is a pM-class
planet.  This dichotomy makes these two planets a particularly interesting 
pair for comparison.

Given these developments, there is a pressing need for 
3D calculations of the atmospheric circulation on hot Jupiters.  Several
groups have investigated a range of 2D and 3D models 
\citep[]{showman-guillot-2002, cho-etal-2003, cho-etal-2008, 
cooper-showman-2005, cooper-showman-2006, showman-etal-2008a,
showman-etal-2008b, langton-laughlin-2007, langton-laughlin-2008, 
dobbs-dixon-lin-2008}. However, to date, all published models 
have adopted severe approximations to the radiative transfer
or excluded radiative heating/cooling entirely.  
While such simplified
approaches are invaluable for investigating the underlying dynamics, 
a detailed attempt to explain wavelength-dependent
photometry and light curves must include a realistic coupling of the
radiative transfer to the dynamics.  

Here, we present new numerical simulations that couple a realistic
representation of non-gray cloud-free radiative transfer to the dynamics.
Surprisingly, coupling of 3D dynamics to nongray radiative transfer
has never previously been done for any giant planet, not even Jupiter, 
Saturn, Uranus, and Neptune.
This is the first 3D dynamical model for any giant planet --- 
inside our Solar System or out ---
to do so.  We dub our model the Substellar and Planetary Atmospheric 
Radiation and Circulation model, or SPARC model, and to honor
its dynamical heritage usually refer to it as SPARC/MITgcm. 
\S 2 presents the model. \S 3 describes the basic circulation regime and
resulting light curves and spectra for HD 189733b.
\S 4 describes the effect of a nonsynchronous rotation rate,
\S 5 considers HD 209458b, and \S 6 concludes.

\section{Model}

\begin{figure}
\vskip 10pt
\includegraphics[scale=0.4, angle=0]{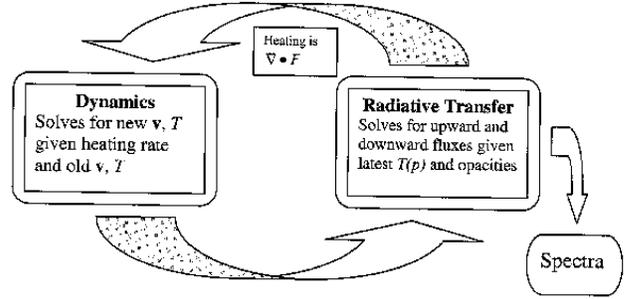}
\caption{Coupled interrelationship between dynamics and radiation in 
our general circulation model (GCM), the SPARC/MITgcm.  
Spectra are calculated offline from GCM
output.}
\label{gcm}
\end{figure}

In the absence of dynamics, radiative transfer drives temperatures toward
radiative equilibrium, which for a tidally locked irradiated planet means
hot on the dayside and cold on the nightside.  These thermal contrasts produce
horizontal pressure gradients that generate winds, which drive the atmosphere
away from radiative equilibrium.  Because of this dynamical response, the
radiative heating/cooling rate is nonzero, and it is this heating/cooling
rate that drives the dynamics.  In our previous work, we calculated the heating
rate using a simplified Newtonian relaxation scheme \citep{showman-guillot-2002,
cooper-showman-2005, cooper-showman-2006, showman-etal-2008a}, but here we 
instead self-consistently
calculate the heating rate from the radiative transfer.  Figure~\ref{gcm}
illustrates the coupling between dynamics and radiation in SPARC/MITgcm.

\subsection{Dynamics}

We solve the global, 3D primitive equations in spherical geometry using
the MITgcm \citep{adcroft-etal-2004}, which is a state-of-the-art atmosphere
and ocean circulation model maintained at MIT.  The primitive equations
are the standard equations for atmospheric flows in statically stable
atmospheres when the horizontal dimensions greatly exceed the 
vertical dimensions.   For HD 189733b and HD 209458b, with
horizontal dimensions of $10^7$--$10^8\m$ and scale heights of 
$\sim200\km$ and $500\km$, respectively, we expect aspect ratios
of $\sim20$--500.  \citet{showman-etal-2008a, showman-etal-2008b} 
provide a more detailed
discussion.  The horizontal momentum, vertical momentum, continuity,
and thermodynamic energy equations are
\begin{equation}
{d{\bf v}\over dt}= -\nabla \Phi - f {\bf k}\times {\bf v} + {\cal D}_{\bf v}
\label{momentum}
\end{equation}
\begin{equation}
{\partial \Phi\over \partial p}=-{1\over\rho}
\label{hydrostatic}
\end{equation}
\begin{equation}
\nabla \cdot {\bf v} + {\partial \omega\over\partial p}=0
\label{continuity}
\end{equation}
\begin{equation}
{d T\over dt} = {q\over c_p} + {\omega \over \rho c_p} + {\cal D}_T
\label{energy}
\end{equation}
where ${\bf v}$ is the horizontal velocity on constant-pressure surfaces, 
$\omega\equiv dp/dt$ is the vertical velocity in pressure coordinates, 
$\Phi$ is the gravitational potential on constant-pressure surfaces, 
$f\equiv 2\Omega\sin\phi$ is the
Coriolis parameter, $\Omega$ is the planetary rotation rate
($2\pi$ over the rotation period), ${\bf k}$ is the
local vertical unit vector, $q$ is the thermodynamic heating rate
($\W\kg^{-1}$), and $T$, $\rho$, and $c_p$ are the
temperature, density, and specific heat at constant pressure.
$\nabla$ is the horizontal gradient
evaluated on constant-pressure surfaces, and $d/dt=\partial/\partial t
+ {\bf v}\cdot \nabla + \omega \partial/\partial p$ is the material
derivative.  Curvature terms are included in ${\bf v}\cdot\nabla {\bf v}$.
Equation~\ref{energy} is actually solved in an alternate form,
\begin{equation}
{d\theta\over dt}={\theta\over T}{q \over c_p} + {\cal D}_{\theta}
\label{energy2}
\end{equation}
where $\theta=T(p/p_0)^{\kappa}$ is the potential temperature
(a measure of entropy), $\kappa$ is the ratio of gas constant to
specific heat at constant pressure, and $p_0$ is a reference
pressure (here chosen to be 1 bar, but note that the dynamics
are independent of the choice of $p_0$).
The dependent variables ${\bf v}$, $\omega$, $\Phi$,
$\rho$, $\theta$, and $T$ are functions of longitude $\lambda$, 
latitude $\phi$, pressure $p$, and time $t$.  
In the above equations, the quantities ${\cal D}_{\bf v}$, 
${\cal D}_T$, and ${\cal D}_{\theta}$ represent additional source/sink 
terms beyond those described explicitly in the equations (see below).

The MITgcm has been widely used in the
atmospheric and ocean-science communities and has been successfully
benchmarked against standard test cases \citep{held-suarez-1994}.
The model, with simplified forcing, has also shown success 
in reproducing the global circulations of giant planets in our Solar System
\citep{lian-showman-2008a}.

\begin{figure}
\vskip 10pt
\includegraphics[scale=0.45, angle=0]{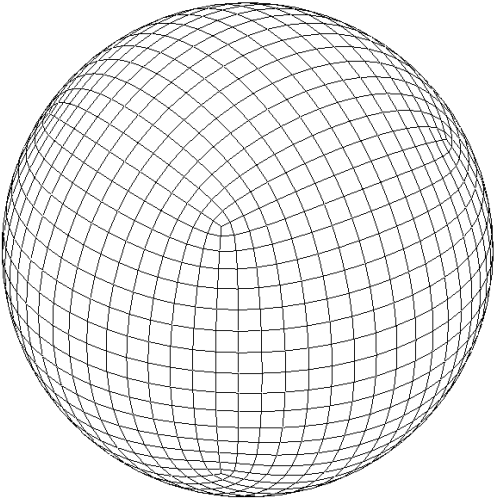}
\caption{
Cubed-sphere grid at a resolution of C16 (i.e., $16\times16$
elements per cube face), roughly equivalent to a global resolution
of $64\times32$ in longitude and latitude.
}
\label{cubed-sphere}
\end{figure}

The MITgcm solves the equations on a staggered Arakawa C grid 
\citep{arakawa-lamb-1977} using a finite-volume discretization.
Two coordinate systems are supported: the standard longitude/latitude coordinate
system and a curvilinear coordinate system called the ``cubed
sphere,'' shown in Fig.~\ref{cubed-sphere}.  In the longitude/latitude
system, the east-west (zonal) grid spacing converges to
zero near the poles.  This convergence implies, via
the Courant-Fredricks-Levy (CFL) constraint, that the timestep
must approach zero to maintain numerical stability.  This difficulty
can be surmounted using a polar filter that smooths the east-west
variability of the dynamical fields near the poles, which increases
the effective grid size and allows use of a finite timestep.  In contrast,
the cubed-sphere grid lacks the singularities at the poles and thereby
sidesteps this problem.  The price is a non-orthogonal grid
and the existence of eight special 
``corner points'' (three of which can be seen in Fig.~\ref{cubed-sphere}). 
Unlike the situation at the poles in a longitude/latitude grid, however,
the coordinate lines in the cubed-sphere grid remain well
separated until very close to the corner points; depending on resolution,
the size of finite-volume elements abutting the corners is typically
one-half to one-third of that away from the corners.  The upshot is
that one can typically use a longer timestep with the cubed-sphere 
grid than a longitude/latitude grid at comparable resolution.  The
simulations presented here adopt the cubed-sphere grid.

The top boundary condition is zero pressure and the bottom boundary
condition is an impermeable surface, which we place far below the
region of interest.   Both boundaries are free slip in horizontal
velocity.   We explore horizontal resolutions of C16,
C32, and C64 (implying that each of the six ``cube faces'' has
a resolution of $16\times16$, $32\times32$, and $64\times64$
finite-volume elements, respectively); these roughly translate 
to global resolutions of $64\times32$, $128\times64$, and $256\times128$ 
in longitude and latitude.   For a model with $N_L$ vertical layers,
the bottom $N_L-1$ layers have interfaces that are evenly spaced in log
pressure between $p_{\rm top}$ and the basal pressure; the top
layer extends from a pressure of zero to $p_{\rm top}$. In most simulations, 
we place the bottom boundary at 200 bars, safely below the active meteorology 
near the photosphere.   We generally place 
$p_{\rm top}$ at 0.2 mbar or $2\,\mu$bar, using 40 or 53 layers,
respectively; in both cases, this resolves each pressure scale height 
with 2.9 layers.

We adopt the third-order Adams-Bashforth timestepping scheme 
\citep{durran-1991}, which has favorable properties over the more commonly
used leapfrog scheme.   To maintain numerical stability, we apply a 
fourth-order Shapiro filter to the time derivatives of ${\bf v}$ and 
$\theta$ at each timestep.  This source/sink, which is represented by
the terms ${\cal D}_{\bf v}$ and ${\cal D}_{\theta}$ in the governing
equations, is analogous to a hyperviscosity: it horizontally smooths 
grid-scale variations while leaving long-wavelength structures relatively 
unaffected.

CFL constraints limit us to a 
short dynamical timestep;
for our C16, C32, and C64 simulations, we generally use $50\sec$, 25 
or $15\sec$, and $6\sec$, respectively.

In our nominal simulations, we assume the planets rotate synchronously
with their orbital periods, implying rotation periods of 2.2 days
($\Omega=3.278\times10^{-5}\,{\rm s}^{-1}$) for HD 189733b and 3.5 days
($\Omega=2.078\times10^{-5}\,{\rm s}^{-1}$) for HD 209458b.  Nevertheless,
because deviations from synchronous rotation are possible
\citep[e.g.][]{showman-guillot-2002}, we also ran some simulations
with differing rotation periods. We adopt gravity and planetary radius
of $21.4\m\,{\rm s}^{-2}$ and $8.2396\times10^7\,$m for HD 189733b and 
$9.36\m\,{\rm s}^{-2}$ and $9.437\times10^7\,$m for HD 209458b.  The equation
of state is ideal gas.  We fix $c_p=1.3\times10^4\J\kg^{-1}\,{\rm K}^{-1}$ and 
use $\kappa=2/7$, appropriate to a predominantly hydrogen atmosphere, 
for both planets.  In most cases the initial condition contains no
winds and adopts an initial temperature profile from a 1D planetwide-average
radiative-equilibrium calculation.

As discussed by \citet{showman-etal-2008b} and \citet{goodman-2009}, the
mean winds that develop in an atmosphere depend in principle
not only on forcing but also on damping processes that remove kinetic 
and/or potential energy.  For solar-system planets, kinetic-energy 
loss occurs via turbulence, wave breaking, and friction against the 
surface (if any).  In the deep interior of gas giants ($p\gtrsim 10^5\,$
bars), Ohmic dissipation may also provide an important source of
drag \citep{kirk-stevenson-1987, liu-etal-2008}, although it is 
unclear whether the atmospheric circulation on hot Jupiters 
couples to such deep regions.  Rigorously representing such frictional 
effects in numerical models is difficult, however; the turbulence associated 
with small-scale shear instabilities and wave breaking often has
length scales much smaller (by up to several orders of magnitude) 
than can easily be resolved in global 3D numerical models.   
In Earth GCMs, these dissipative processes are generally parameterized
by adding to the equations empirical damping terms (e.g.,
a vertical diffusion to represent kinetic-energy losses by
subgrid-scale shear instabilities and waves).  A difficulty is that 
such prescriptions, while physically motivated, are generally non-rigorous 
and the extent to which they can be extrapolated to hot Jupiters is 
unclear. Moreover, simulations of the cloud-level circulation on Jupiter, 
Saturn, Uranus, and Neptune that lack frictional drag parameterizations have 
nevertheless shown success in reproducing the qualitative features of the
observed flow on these planets \citep[e.g.,][]{scott-polvani-2007, 
scott-polvani-2008, showman-2007, lian-showman-2008a, lian-showman-2009}.
Therefore, while recognizing the possible importance that
frictional processes could play for hot Jupiters, for the present study 
we do not include frictional damping terms in the momentum equation 
(Eq.~\ref{momentum}) other than the Shapiro filter, which causes a diffusive 
damping of small-scale wind structures.  
Once our simulations have spun up, this implies that,
except for energy losses due to the Shapiro filter, the globally integrated 
rate of production of available potential energy 
\citep[APE; see][chapter 14]{peixoto-oort-1992} and its net conversion to kinetic 
energy (KE) become modest, with the potential-to-kinetic energy 
conversion in some regions of the 
atmosphere being counteracted by the kinetic-to-potential energy
conversion in other regions.\footnote{For example, our simulations
develop a broad eastward equatorial jet; near IR photosphere levels,
where this jet transports air from day-to-night, the air generally flows 
in the same direction as the horizontal pressure-gradient force
(i.e. ${\bf v}\cdot{\nabla\Phi}<0$), so potential energy is converted
to kinetic energy.  However, where the jet transports air from night-to-day,
such a jet tends to flows against the horizontal pressure-gradient force
(${\bf v}\cdot{\nabla\Phi}>0$), so kinetic energy is converted
to potential energy.  In a global mean, there is a large degree of
cancellation between these competing effects.}  
In a future study, we will explore the influence 
that plausible frictional parameterizations exert on the wind speeds
and circulation geometry.

\citet{goodman-2009} suggested that the heating associated with
frictional dissipation should be included in the thermodynamic
energy equation.  Our study, like previous published hot-Jupiter studies,
does not include this effect; this is equivalent to assuming
that any frictional heating is small compared to the radiative
heating.  This is typically a reasonable assumption for atmospheres:
the globally averaged rate of kinetic-energy production
(and hence the rate at which that kinetic energy can be dissipated
into thermal energy) is controlled by the thermodynamic efficiency 
of the atmospheric heat engine, which is typically much less than 100\%,
at least for atmospheres in the Solar System.  In this situation
the frictional heating is much less than the radiative heating.
On Earth, for example, 
the globally averaged rate of frictional dissipation by the large-scale 
circulation is $2\rm\,W\,m^{-2}$, which is only a few percent of the 
latitude-dependent net radiative heating/cooling rate 
\citep{peixoto-oort-1992}.  This implies that the frictional heating is 
only a small perturbation to the total heating rate. 
Because giant planets lack surfaces (which is the primary source of
friction on Earth) is it plausible that the frictional damping,
and the heating it causes, represents an even smaller effect on gas 
giants than terrestrial planets.  Note that, to date, frictional heating
has not been included in cloud-level atmosphere simulations of Jupiter, Saturn,
Uranus, and Neptune.  For hot Jupiters, given the considerable uncertainty
arising from other factors (e.g., uncertainties in opacities, composition,  
possible presence of clouds/hazes, and possible non-synchronous rotation,
all of which influence the circulation), it seems reasonable
to neglect this effect for the present.

\subsection{Radiative transfer calculation}

We coupled the MITgcm to the plane-parallel radiative-transfer 
code of \citet{marley-mckay-1999}, which is a state-of-the-art
model that has been extensively used in 1D investigations of Titan, 
the giant planets, brown dwarfs, and exoplanets
\citep[e.g.][]{mckay-etal-1989, marley-etal-1996, burrows-etal-1997, 
marley-etal-1999, marley-etal-2002, 
fortney-etal-2005, fortney-etal-2006a, fortney-etal-2006b, 
fortney-marley-2007, fortney-etal-2008, saumon-marley-2008}.  
We here use the code in its two-stream formulation, allowing
treatment of the upward and downward radiative fluxes versus wavelength and
height throughout the 3D grid.  Multiple scattering is properly accounted for.

\subsubsection{Opacities}

The code treats opacities using the correlated-k method 
\citep{goody-yung-1989}, which is ideal for use in GCMs because it is 
fast and accurate.  In this approach opacities are first computed exactly 
at millions of individual frequency points at over 700 pressure-temperature 
($p$--$T$) points, accounting for the influence of all molecular and atomic
lines in our opacity database \citep{freedman-etal-2008}.  Line broadening 
is computed for each line of each species at each pressure-temperature 
point.  The summed opacities at each $p$--$T$ point
thus account for both the relative abundances of each absorbing species 
as well as the line shapes appropriate for the particular physical 
conditions. The spectrum of opacities is then divided into 30 frequency 
bins\footnote{In recent 1D applications of our code we typically use 196 
intervals \citep[e.g.][]{fortney-etal-2005, fortney-etal-2006a, 
fortney-etal-2006b, fortney-etal-2008}.  Extensive testing has revealed  
that, for a specified  $T(p)$ profile, the net radiative flux 
calculated using 30 bins differs less than $1\%$ from that calculated 
using 196 and is substantially faster in the GCM environment.} 
as listed in Table~\ref{bins-table}.

The opacities at each of the several tens of thousands of frequency point 
within each frequency bin (Table~\ref{bins-table}) 
are then sorted by strength and 
described statistically thus enabling gaussian integration
to be used to solve for the radiative fluxes through the model eight times, 
once for each probability interval within each spectral bin.  The model 
flux within each of the 30 frequency bins is the sum of the
flux calculations weighted by the relative contribution of each statistical 
opacity interval.  By using two intervals (one for the opacities weaker 
than 95\% of the highest value and one for the
strongest 5\%), each with 4 gauss points (for 8 total gauss points), 
we resolve both the important strongest lines as well as the background 
continuum  within each bin.  This k-coefficient approach has been widely 
used in planetary atmospheres calculations for over two decades and
is frequently employed in GCM calculations.  The method is subject to 
significant errors only when the distribution of opacity within a single 
spectral bin varies substantially with height in the atmosphere as might 
happen if one absorber is replaced by another with a very different opacity 
shape (for example a strong red slope within a bin replaced by a strong 
blue opacity slope {\em within that same bin}).  In practice such concerns 
are mitigated by careful choice of the opacity intervals.  To compute 
the opacity at arbitrary $p$--$T$ points we interpolate within our grid of 
k-coefficients.  

\begin{figure}
\vskip 10pt
\includegraphics[scale=0.45, angle=0]{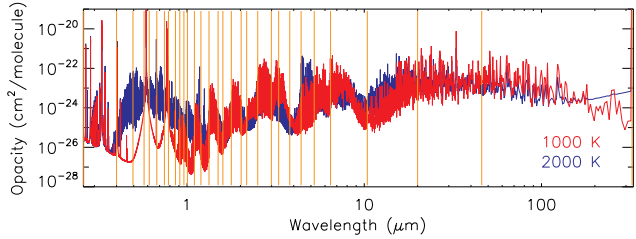}
\caption{Opacity (in $\rm cm^2\, molecule^{-1}$) versus wavelength
at $p=1\rm\,bar$ and $T=1000\rm\,K$ (red) and $2000\rm\,K$ (blue)
assuming solar composition (modified for rainout where appropriate)
including TiO and VO.  In the mid-IR, much of the structure results
from H$_2$O, CO, and (at 1000 K) CH$_4$; the two prominent Na lines are 
evident in the visible in the 1000-K case.  Significant additional
visible-wavelength opacity occurs in the 2000-K case due to
absorption by TiO and VO.  Wavelength boundaries for the 30 opacity bins in
our GCM runs are shown with vertical orange lines and listed numerically
in Table~\ref{bins-table}. 
}
\label{bins-figure}
\end{figure}

We chose the wavelength boundaries of our bins in an attempt optimize a 
number of competing needs.  
We wish to have highest spectral resolution (narrowest bins) in the optical 
at the peak of the incident flux (to best resolve absorption of incident 
energy) and in the near-infrared to similarly resolve the peak of the 
emergent flux.  In addition, distinct bins should resolve important opacity 
peaks and valleys so that there is a relatively flat opacity distribution 
within each bin.  This minimizes the dynamic range that must be covered by 
the 8 k-coefficient Gauss points within each bin.
Examples include the water opacity windows in the 
near-infrared, the CO bandhead at $2.3\,\rm \mu m$, and TiO in the optical.  
We also wish to minimize the total number of bins while still covering the 
entire spectral range from the UV to the thermal infrared.  Our choice of 
bins is listed in Table~\ref{bins-table} and depicted graphically
in Fig.~\ref{bins-figure}. 

\begin{deluxetable}{cc}
\tablecolumns{2}
\tablewidth{0pc}
\tablecaption{Opacity Bins}
\tablehead{
\colhead{wavelength ($\mu$m)} & \colhead{wavelength ($\mu$m)}}
\startdata
   324.68 &  46.00\\
    46.00 &  20.00\\
    20.00 &  10.40\\
    10.40 & 6.452\\
   6.452 & 5.220\\
   5.220 & 4.400\\
   4.400 & 3.800\\
   3.800 & 3.288\\
   3.288 & 2.989\\
   2.989 & 2.505\\
   2.505 & 2.170\\
   2.170 & 2.020\\
   2.020 & 1.777\\
   1.777 & 1.593\\
   1.593 & 1.497\\
   1.497 & 1.330\\
   1.330 & 1.197\\
   1.197 & 1.100\\
   1.100 & 1.005\\
   1.005 & 0.960\\
   0.960 & 0.910\\
   0.910 & 0.860\\
   0.860 & 0.785\\
   0.785 & 0.745\\
   0.745 & 0.675\\
   0.675 & 0.612\\
   0.612 & 0.572\\
   0.572 & 0.495\\
   0.495 & 0.400\\
   0.400 & 0.261\\
\enddata
\tablecomments{Opacity bins used in the radiative-transfer calculation.
Each row lists the bounding wavelengths for one bin.}
\label{bins-table}
\end{deluxetable}

The opacities themselves are computed 
from a comprehensive opacity database \citep{freedman-etal-2008}, 
calculated assuming local thermodynamic and chemical equilibrium 
\citep{lodders-fegley-2002, lodders-fegley-2006} as a function of
$p$, $T$, and wavelength.   We perform simulations with 
opacities corresponding to 1, 5, and 10 times solar abundances 
(appropriately modified for rainout of condensed species whenever
condensation should occur). For HD 189733b, the inferred water abundance 
\citep{tinetti-etal-2007b, swain-etal-2008} 
is consistent with near-solar metallicity for both water and 
carbon \citep{showman-2008}.  For HD 209458b, \citet{barman-2007}'s 
fits to transit data suggest abundances not exceeding 
a few times solar, while \citet{sing-etal-2008}'s  fits to transit 
data suggest two times solar sodium abundance.  

TiO and VO deserve special mention.  For our simulations of HD 189733b, 
we use opacity databases that exclude TiO and VO,
consistent with secondary-eclipse data \citep{charbonneau-etal-2008}, 
1D radiative-transfer studies  \citep{barman-2008, knutson-etal-2009a}, 
and evolution models suggesting that TiO and
VO should sequester deeper than $\sim10$--$100\,$bars where solid clouds
incorporating Ti and V form \citep[e.g.][]{fortney-etal-2006a}.  
For HD 209458b, however, the detection of a 
hot stratosphere \citep{knutson-etal-2008a,burrows-etal-2007b}, possibly
associated with TiO and VO \citep{hubeny-etal-2003, fortney-etal-2008} 
suggest that gaseous TiO and VO may exist in this planet's atmosphere.
For HD 209458b, therefore, we run some cases that include gaseous TiO and VO
and other cases that exclude them.

For simplicity, we neglect cloud opacities.  This appears to be reasonable
for HD 189733b, at least for emitted thermal radiation, since
absorption features observed in transit spectra longward of $\sim1.5\,\mu$m 
\citep{tinetti-etal-2007b, swain-etal-2008, sing-etal-2008} indicate
that gas rather than particles dominates the opacity along the slant transit 
path.  The situation is less clear at 
visible wavelengths \citep{pont-etal-2008, 
redfield-etal-2008}, but submicron-sized haze particles are a possibility.
In any event, because the {\it slant} optical depth exceeds the 
{\it normal} optical depth by a factor of $\sim50$ \citep{fortney-2005}, 
the data are nevertheless consistent with normal cloud optical
depths $\ll 1$.   Fewer constraints exist for HD 209458b.  Some authors
have suggested silicate absorption features in emission spectra
\citep{richardson-etal-2007}, but debate exists \citep{swain-etal-2007}.

\subsubsection{Flux Calculation}

We solve for the net flux through each level of a specified local atmospheric 
structure using two separate techniques: one for the incident stellar 
and one for the emitted planetary thermal radiation.  This subdivision
is common in planetary-atmosphere studies since the incident radiation is 
dominated by the direct stellar beam accompanied by a weaker but more 
uniform scattered flux while the thermal emitted radiation is comparatively 
much more uniform over a hemisphere, thus suggesting separate flux solution 
methods.  For solar-system atmospheres the solar and planetary fluxes 
usually occupy distinct spectral
intervals.  However for hot extrasolar planets the incident and emitted 
fluxes substantially overlap near $1\,\rm \mu m$ and thus the
net flux within a layer is the sum of the net stellar and thermal fluxes, 
which are computed by differing techniques within the same spectral interval.

For the incident stellar flux we follow \cite{mckay-etal-1989} and 
\cite{marley-mckay-1999} and employ the delta-discrete ordinates 
method (see Appendix I of \cite{mckay-etal-1989}).  
We adopt stellar spectra from Kurucz 
(see http://hurucz.harvard.edu/stars.html)
and assume that HD 189733b and HD 209458b have distances of 0.0313
and 0.046 AU from their stars, respectively.  Obliquities
and orbital eccentricities are assumed zero.  Stellar fluxes are computed 
only in those bins lying shortwards of $6\,\rm \mu m$.

For the thermal flux we follow the approach used in our brown dwarf work 
and employ the ``two-stream source function'' technique
of \citet{toon-etal-1989} which was specifically developed for 
rapid, accurate computation of radiative heating rates
in inhomogeneous atmospheres.  This approach is well suited for atmospheres 
with scattering cloud layers, which we will eventually include.  The present
implementation does include Rayleigh scattering, although this is generally 
unimportant for most of the thermal bins.  The
accuracy of this technique for various limiting cases is discussed in detail 
by \cite{toon-etal-1989}.  We model thermal radiation from the planet from 
0.26--$325\,\mu$m.  Extensive experience with this technique employed with 
k-coefficient opacities in the context of brown dwarf atmospheres has 
shown that integrated fluxes usually agree to much better than 0.1\% with 
fluxes computed using detailed line-by-line calculations.

In most 1D radiative-transfer calculations, an iteration is performed to
drive the solution into radiative equilibrium.  Here, the 3D solution is
not in radiative equilibrium, and no iteration is needed.  Instead,
at each timestep, we pass a driver subroutine
the most recent 3D temperature field, consisting of numerous individual
$T(p)$ columns, one per element in the horizontal grid.  We loop over all
these columns and, for each, 
we call the radiative-transfer solver once (without iteration)
to determine the upward and downward fluxes versus frequency  
and pressure throughout that column.  For each column, we include
the appropriate incident starlight: columns on the nightside emit
radiation but receive no starlight, 
while columns on the dayside receive starlight
along an angle $\theta$ from the local vertical.  The cosine of this angle,
$\mu$, varies from 1 at the substellar point to 0 at the terminator.  
Columns closer to the substellar point thus receive greater stellar flux,
and this flux penetrates to greater pressures because of the shorter
atmospheric path lengths.

\subsection{Coupling dynamics and radiative-transfer}

Once we calculate the wavelength-dependent radiative fluxes, we sum them to
obtain the net vertical flux everywhere over the 3D grid at that timestep.  
We then obtain the thermodynamic heating rate $q$ (Eq.~\ref{energy}) 
as follows.  If an atmospheric layer absorbs more 
(less) radiation than it emits, the layer experiences heating (cooling).  
 The thermodynamic heating rate per volume ($\W\m^{-3}$)
is thus simply $-\partial F/\partial z$, where $z$ is height and $F$ is
the net, wavelength-integrated flux.  Expressed in pressure coordinates 
using hydrostatic balance, the heating rate per mass, $q$, becomes
\begin{equation}
q= g {\partial F\over\partial p}
\label{heating}
\end{equation}
This is then used to drive the dynamics via Eq.~\ref{energy2}
(see Fig.~\ref{gcm}).
Note that this expression neglects the heating contribution 
due to the horizontal divergence of the horizontal radiative flux, 
which is at least a factor of $\sim30$ less than that in Eq.~\ref{heating} 
for conditions relevant to hot Jupiters.  

We use the same vertical gridding for the dynamics and radiative-transfer
calculations. 
In the SPARC/MITgcm, the vertical grid is staggered such that $T$, $\theta$, 
and horizontal wind are defined within the layers, while certain other 
quantities (such as vertical velocity) are defined at the interfaces 
between layers.  Because $q$ is used to update $\theta$ (Eq.~\ref{energy2}),
$q$ should be computed within the layers so that it has the same vertical
positioning as $\theta$.  To calculate $q$ we thus evaluate 
$\partial F/\partial p$ by finite differencing the fluxes and pressures 
between the {\it interfaces} that over- and underlie a given layer.  
This maximizes accuracy while positioning $\partial F/\partial p$ within 
the layer.  Note that we use the same layer spacing in the dynamics
and radiative transfer calculations; 1D radiative transfer tests performed
offline suggest that this vertical gridding is sufficient to resolve
the radiative transfer.

As a test, we ran a suite of simulations where we
turned off the dynamics.  In these cases, the temperature profiles 
relaxed toward the expected radiative equilibrium (as defined by
solutions generated with the radiative-equilibrium 1D version of the 
radiation code) for the specified value of $\mu$.  
This demonstrates that the calculation of the thermodynamic
heating rate (Eq.~\ref{heating}) functions properly.

In most SPARC/MITgcm simulations, 
we update the radiative fluxes less frequently 
than the dynamical timestep.  This is standard practice in GCMs and 
helps maintain computational efficiency. For our simulations without 
TiO and VO, we found that a radiative step of 500--$1000\sec$ was
generally sufficient.   For simulations with TiO and VO, however, a much
shorter radiative timestep of $100\sec$ or less was necessary to 
maintain numerical stability.  For computational expediency, in
some simulations, particularly our high-resolution cases, we
perform radiative transfer only on every other grid column and interpolate
the fluxes and heating rates in between.

SPARC/MITgcm output includes the emergent fluxes 
versus wavelength and position
everywhere over the globe.  However, because we run the 
GCM with only modest spectral resolution, we generally recalculate 
disk-averaged light curves and spectra offline at higher spectral resolution
using the 3D temperature fields following
the procedure described in \citet{fortney-etal-2006b} and 
\citet{showman-etal-2008a}.  This procedure ensures that limb
darkening is properly treated.

\section{Results: Synchronously rotating HD 189733b}

\subsection{Circulation regime}

\begin{figure}
\vskip 10pt
\includegraphics[scale=0.43, angle=0]{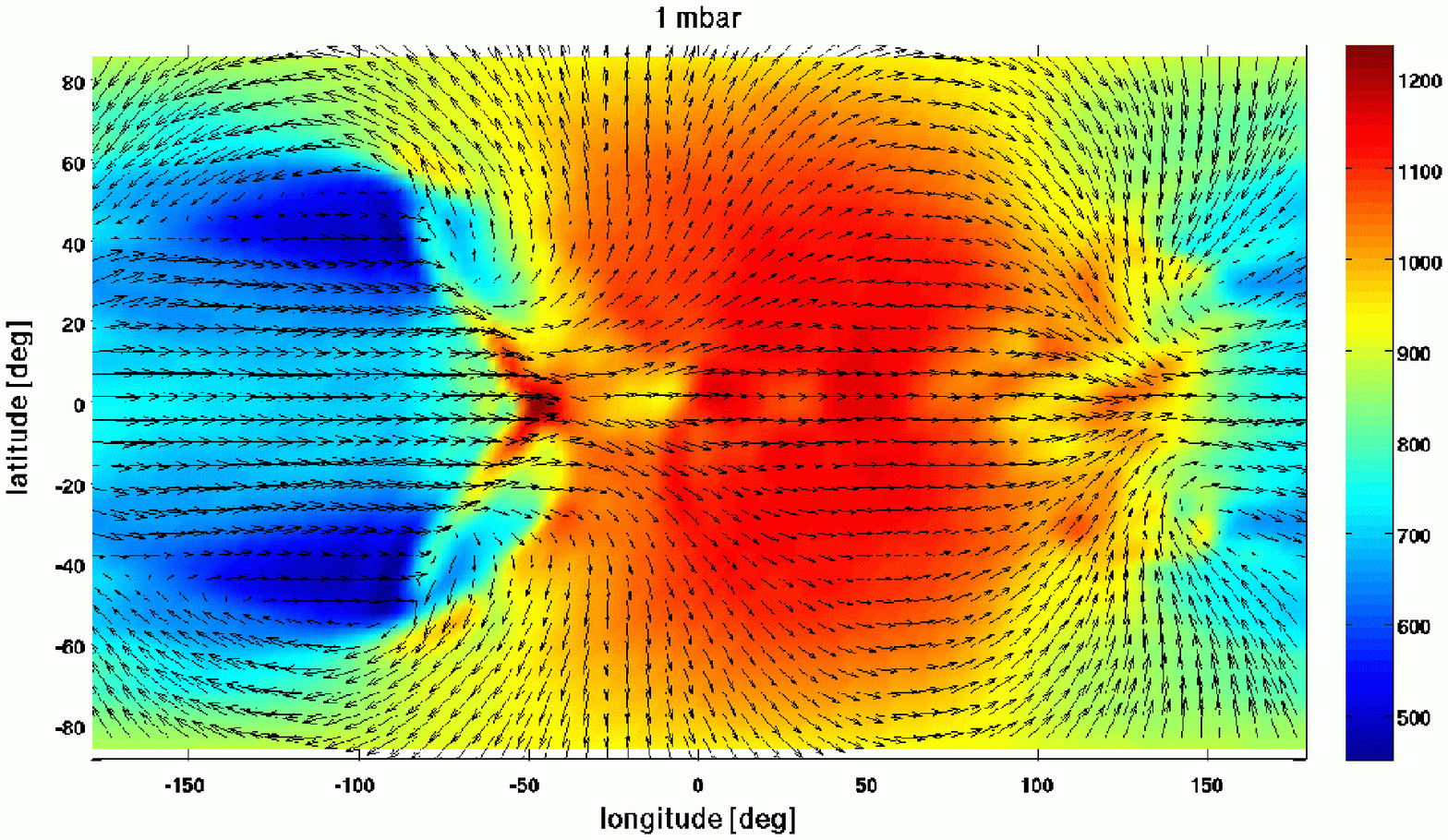}
\includegraphics[scale=0.43, angle=0]{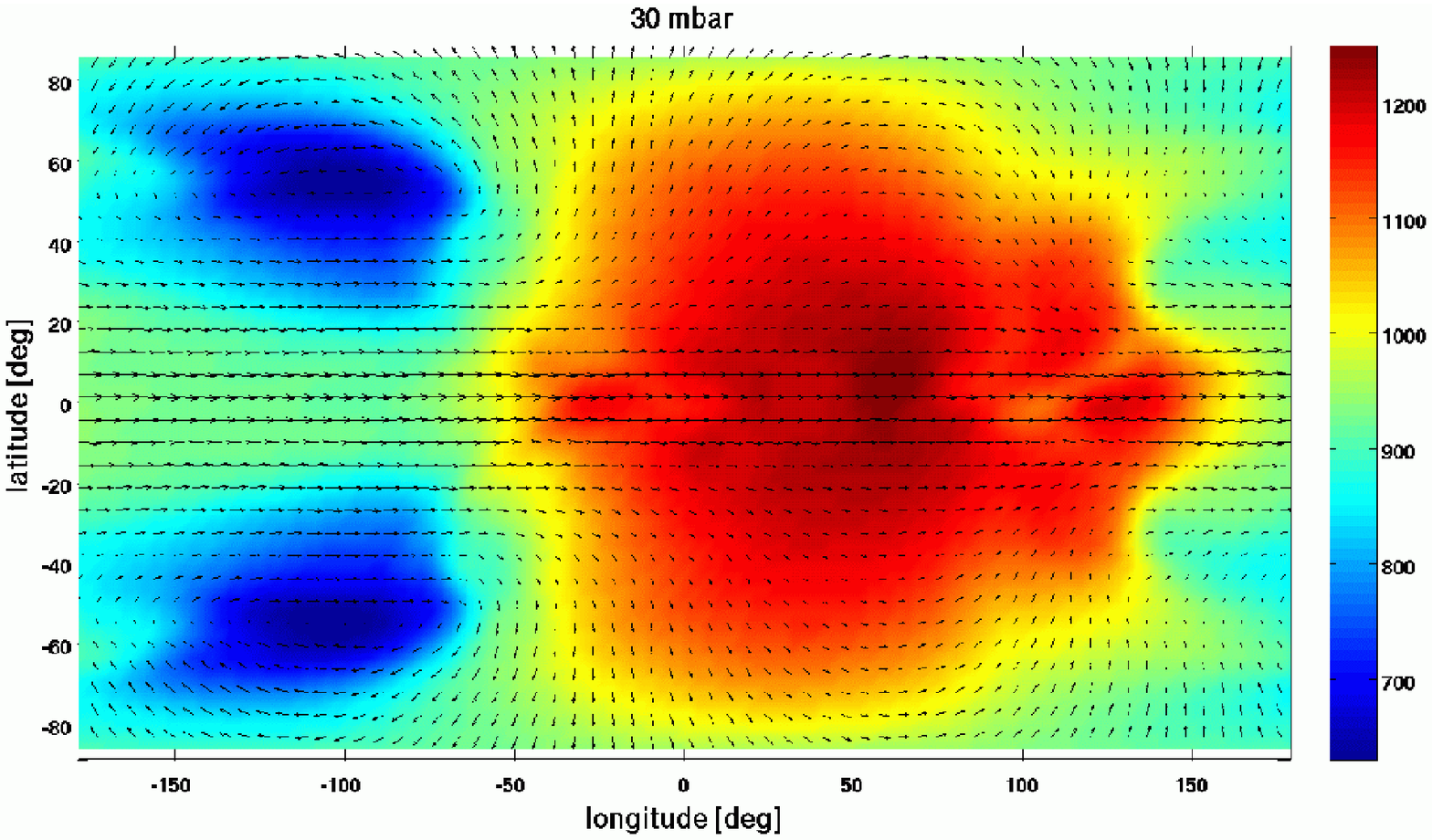}
\includegraphics[scale=0.43, angle=0]{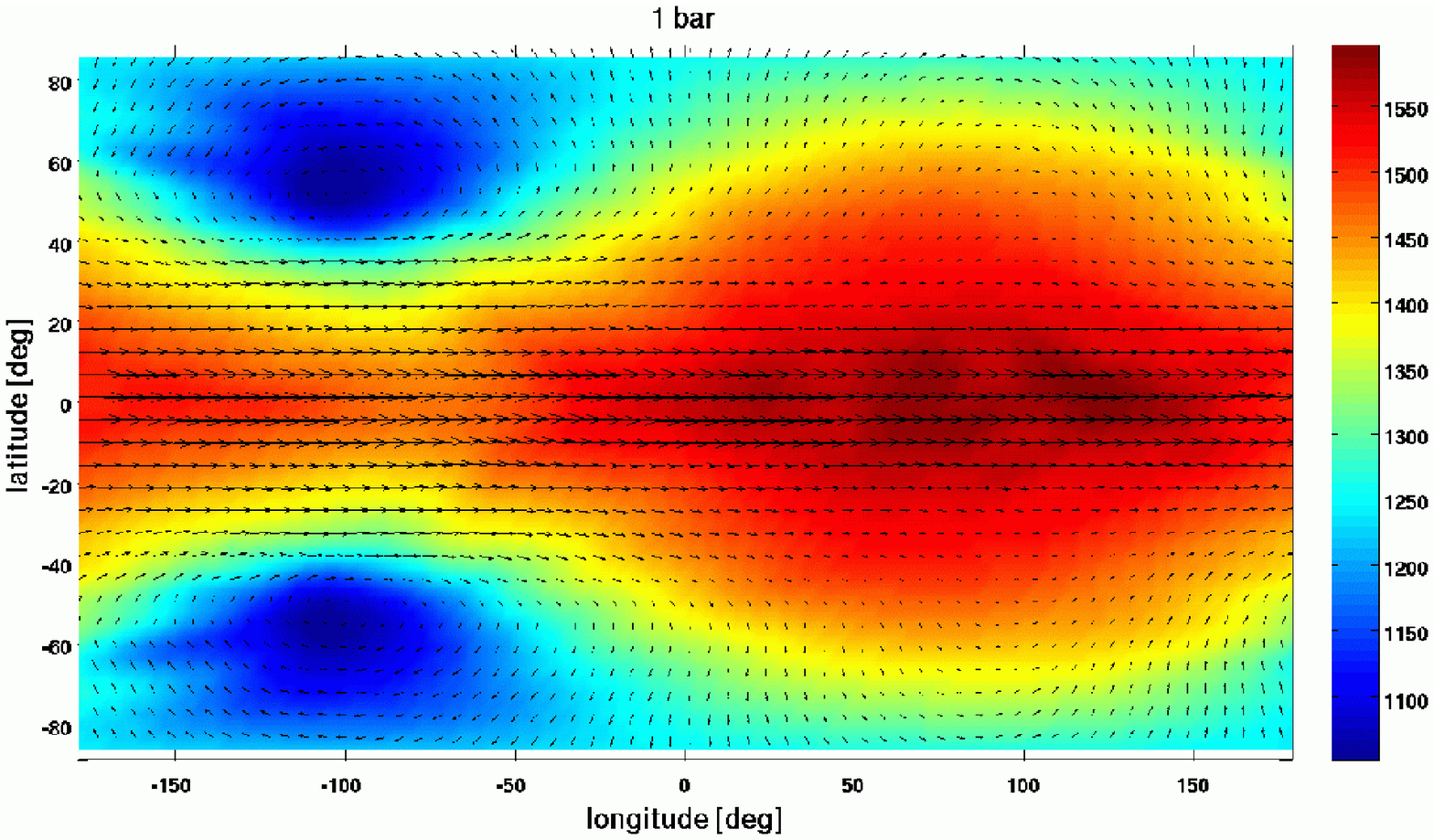}
\caption{Temperature (colorscale, in K) and winds (arrows) for nominal
HD 189733b simulation with solar abundances and no TiO/VO. Panels 
show flow at 1 mbar ({\it top}); 30 mbar, corresponding to an 
approximate photosphere level in the mid-IR ({\it middle}); and
1 bar ({\it bottom}). Resolution
is C32 (roughly equivalent to a global horizontal resolution of $128\times64$
in longitude and latitude) with 40 vertical layers.   
Substellar point is at longitude, latitude $(0^{\circ},0^{\circ})$.
Dayside is region between longitudes $-90^{\circ}$ and $90^{\circ}$.  
Nightside is at longitudes $-180^{\circ}$ to $-90^{\circ}$ and 
$90^{\circ}$ to $180^{\circ}$.
}
\label{hd189-solar}
\end{figure}

\begin{figure}
\vskip 10pt
\includegraphics[scale=0.6, angle=0]{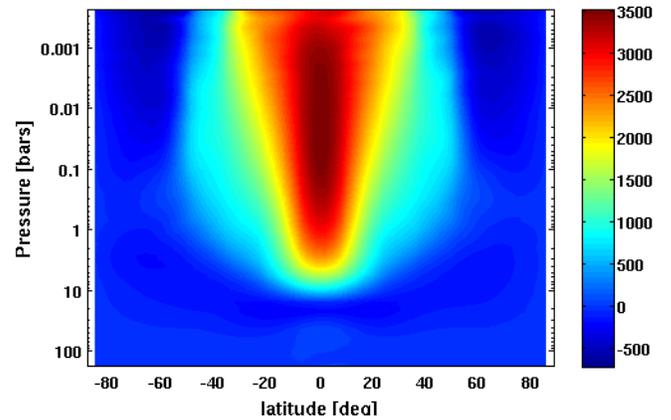}
\caption{Zonal-mean zonal winds for our nominal HD 189733b simulation at
solar abundance.  This is the same simulation as in Fig.~\ref{hd189-solar}.
Scalebar gives speeds in m$\,{\rm s}^{-1}$.}
\label{hd189-ubar}
\end{figure}

In our simulations, the imposed day-night irradiation gradient rapidly
leads to the development of large day-night temperature contrasts and
winds reaching several $\km\sec^{-1}$.   A statistically steady flow
pattern is achieved at pressures $<1\,$bar after less than 100 Earth days
of integration. Figures~\ref{hd189-solar}--\ref{hd189-ubar} illustrate
the behavior for our nominal, synchronously rotating simulation of HD 
189733b using solar opacities without TiO and VO.  Figure~\ref{hd189-solar}
shows the temperature (colorscale) and winds (arrows) over the globe on three
different pressure levels, while Fig.~\ref{hd189-ubar} depicts the
zonally averaged zonal wind\footnote{In atmospheric dynamics, the terms
zonal and meridional wind refer to the east-west and north-south wind,
respectively (with eastward and northward being positive); a zonal average
refers to an average of any quantity in longitude.} versus 
latitude and pressure.

As in our previous simulations with simplified
forcing \citep{showman-guillot-2002, cooper-showman-2005, cooper-showman-2006,
showman-etal-2008a}, the flow becomes dominated by a superrotating (eastward)
equatorial jet extending from latitudes of approximately $30^{\circ}$N to
$30^{\circ}$S.  Vertically, the jet remains coherent from the top of 
the model (0.2 mbar) to the $\sim10$-bar level and reaches peak zonal-mean 
zonal wind speeds of $3.5\,{\rm km}\,{\rm sec}^{-1}$ at 
pressures of 10--100 mbar (Fig.~\ref{hd189-ubar}).  The latitudinal 
width and longitudinal structure of the jet depend strongly on altitude.  
Longitudinal variability in the jet speed 
is small at pressures exceeding $\sim200\,$mbar but increases with altitude
and exceeds $2\,{\rm km}\,{\rm sec}^{-1}$ at the top of the model, with 
the fastest eastward winds occurring in the hemisphere west of the 
substellar point and the slowest winds occurring near and east 
of the substellar point.  The transition between these regions 
narrows with altitude, and at pressures less than a few mbar, the 
equatorial flow appears to pass through a hydraulic jump as it approaches 
the substellar point from the west (e.g., Fig.~\ref{hd189-solar}, 
{\it top panel}), 
leading to a near-discontinuity in the zonal and meridional wind speed 
at longitudes of $-50^{\circ}$ to $-80^{\circ}$ (depending on latitude).
Accompanying this structure are strong downward velocities, and the
adiabatic compression associated with this descent leads to a 
warm chevron-shaped structure in the temperature (Fig.~\ref{hd189-solar},
{\it top panel}).

At higher latitudes (poleward of $\sim45^{\circ}$), the 
zonal-mean zonal wind is westward at pressures $<1\,$bar 
(Fig.~\ref{hd189-ubar}).  Nevertheless, this high-latitude flow exhibits 
complex structure, with westward flow occurring west of the 
substellar point, eastward flow east of the substellar point, and 
substantial day-to-night flow over the poles.  These polar flows 
indicate the importance of performing global simulations
and would not be properly captured if the poles were removed
from the computation domain, as done by some previous authors
\citep[][]{dobbs-dixon-lin-2008}.

On average, the dayside is hotter than the nightside, but the dynamics distorts
the temperature pattern in a complex manner (Fig.~\ref{hd189-solar}).  
Perhaps most importantly, 
the hottest regions do not occur at the substellar longitude; instead,
advection associated with the equatorial jet shifts 
the hottest regions downwind (eastward) of the substellar point by 
$\sim50^{\circ}$.  This point has been previously emphasized
by \citet{showman-guillot-2002}, \citet{cooper-showman-2005}, and
\citet{showman-etal-2008a}.  Moreover, the coldest regions do not 
occur at the equator but instead within two broad
gyres centered at latitudes of $\pm40$--$50^{\circ}$ and longitudes 
$\sim60$--$80^{\circ}$ east of the antistellar point --- a phenomenon
not seen in our previous simulations with simplified forcing
\citep{showman-etal-2008a}.  The horizontal wind speed
is almost zero near the center of these gyres, so air parcels trapped 
there have long residence times on the nightside.  This allows
them to experience a large temperature drop due to radiative cooling.
In contrast, air within the equatorial jet has only a short
residence time (typically $\sim1\,$day) on the nightside because of the 
fast jet speeds, leading to only modest temperature decreases
via radiative cooling.   Interestingly,
the temperature structure shows substantial longitudinal variability even
at pressures as great as $1\,$bar.

As compared to our previous simulations with simplified forcing
\citep{showman-etal-2008a}, our current HD 189733b simulations 
exhibit modest lateral temperature contrasts.  The horizontal 
temperature differences reach $\sim450\,$K at 1 bar and $\sim750\,$K 
at 1 mbar (Fig.~\ref{hd189-solar}).  In contrast, in our previous 
simulations forced by Newtonian heating/cooling \citep{showman-etal-2008a},
the day-night temperature differences reached nearly $900\,$K 
at 100 mbar and $1000\,$K at 10 mbar.  The smaller values in
our present simulations, which can be attributed to our usage
of realistic radiative transfer, have major implications for
light curves and spectra (\S 3.2).

Notice that the global-scale temperature structure exhibits significant
vertical coherency throughout the observable atmosphere 
(Fig.~\ref{hd189-solar}).  Although the detailed structure varies between
levels, the hottest regions lie east of the substellar point throughout, 
with the longitudinal offset of the hottest region varying only modestly 
between pressures of 1 bar and 1 mbar.  Likewise, the 
locations (though not the shape) of the coldest regions also maintain 
coherency across this pressure range (blue regions in Fig.~\ref{hd189-solar}).  
At first glance, this vertical coherency is surprising, because idealized
radiative calculations have shown that the radiative time constant should
vary by orders of magnitude over this pressure range \citep{iro-etal-2005,
fortney-etal-2008, showman-etal-2008a}.  At pressures where the radiative
time constant is comparable to the time for wind to advect across a 
hemisphere, one expects a significant offset of the hottest regions
from the substellar point \citep{showman-guillot-2002}.  However,
at low pressures where the expected radiative time constants are much
shorter than plausible advection times, one expects the temperature patterns
to track the stellar heating, with the hottest region occurring close to
the substellar point \citep{cooper-showman-2005, knutson-etal-2007b}.  
Indeed, precisely this height-dependent pattern is seen in published 3D 
simulations that force the flow with a simplified Newtonian 
heating/cooling scheme, which relaxes the temperature toward the 
radiative-equilibrium temperature profile over the 
expected radiative timescale \citep{showman-etal-2008a, fortney-etal-2006b,
cooper-showman-2005, cooper-showman-2006}. 

So what causes the vertical coherency in our current simulations? 
The simple arguments described above --- in which the temperature 
approaches radiative equilibrium if radiation times are less than 
advection times --- implicitly assume that
the radiative-equilibrium temperature profile can be 
independently defined and that it has a structure reflecting 
that of the insolation, with the greatest radiative-equilibrium
temperature at the substellar point.  However, this argument 
neglects the fact that, in real radiative
transfer, the {\it radiative-equilibrium temperature profile itself} 
depends on the dynamical response and
can involve radiative interactions between different levels.\footnote{A more
rigorous way of stating this is that one cannot define a radiative equilibrium
temperature structure in isolation from the dynamics.  
In that case, the comparison-of-timescales argument fails and 
one must solve the full radiative-dynamical problem, as
we are doing here.}  To illustrate, suppose
the small heating rates at $\sim1\,$bar lead to 
a hot region shifted east of the substellar point
(as seen in Fig.~\ref{hd189-solar}, {\it bottom panel}).    Upwelling 
infrared radiation from these hot, deep regions will warm the entire overlying
atmosphere, leading to a temperature pattern at low pressure that has
similar spatial structure as that at higher pressure.  These effects
were ignored in our previous studies adopting Newtonian cooling
\citep{showman-etal-2008a, fortney-etal-2006b, cooper-showman-2005, 
cooper-showman-2006}, but they are self-consistently included here and 
can explain the vertical coherency in Fig.~\ref{hd189-solar} despite the
large expected vertical variations in radiative time constant.
Nevertheless, our 5 and 10 times solar HD 189733b simulations 
(and especially the HD 209458b simulations with TiO and VO opacity
to be discussed in \S 5)
exhibit less vertical coherency than shown in Fig.~\ref{hd189-solar}.

\begin{figure}
\vskip 10pt
\includegraphics[scale=0.55, angle=0]{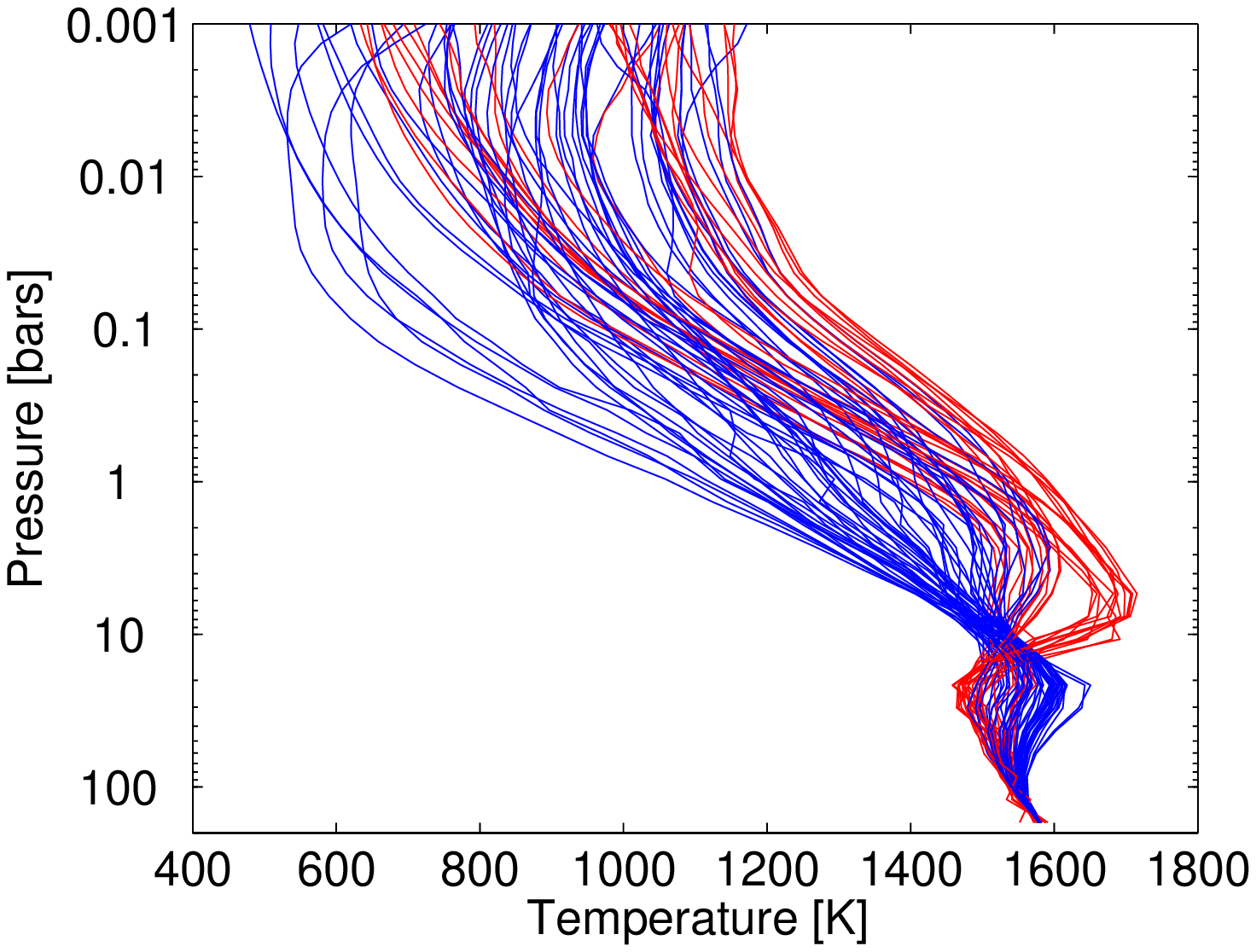}
\includegraphics[scale=0.55, angle=0]{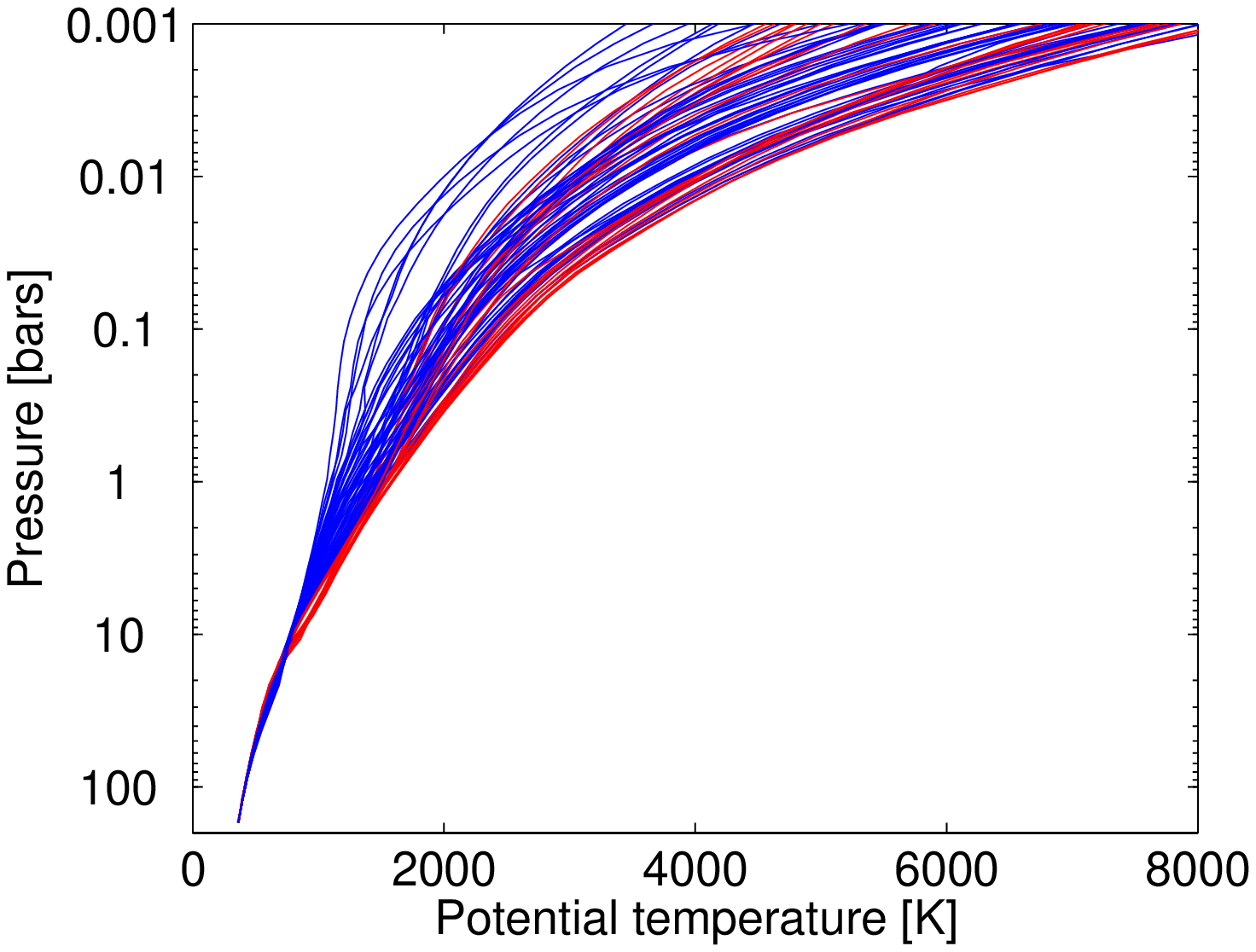}
\caption{A selection of profiles of temperature ({\it top}) and potential temperature
({\it bottom}) versus pressure in our nominal, solar-abundance HD 189733b simulation (the same simulation as in Fig.~\ref{hd189-solar}).
Red (blue) profiles are equatorward (poleward) of $30^{\circ}$ latitude.}
\label{hd189-spaghetti}
\end{figure}

Despite these quantitative differences, our current simulations
lie within the same basic dynamical regime as our previous 3D
simulations driven by Newtonian cooling \citep{showman-etal-2008a, 
cooper-showman-2005, cooper-showman-2006}.  In all these cases,
the flow exhibits a broad, eastward equatorial jet with
westward flow at high latitudes; eastward displacements of the hottest
regions from the substellar point (at least over some range of pressures); 
and a gradual transition from
a banded flow at depth to a less-banded flow aloft.  Moreover, in all these
cases, the flow structures have horizontal lengthscales comparable to 
the planetary radius, consistent with the large Rhines scale and
Rossby deformation radius for these planets \citep{showman-guillot-2002,
menou-etal-2003, showman-etal-2008b}.

Figure~\ref{hd189-spaghetti}, top panel, illustrates the diversity of vertical
temperature profiles that occur. 
Red (blue) profiles lie equatorward (poleward) of $30^{\circ}$ latitude. 
This is for our nominal HD 189733b 
simulation with solar abundance; qualitatively similar patterns occur
for 5 and 10 times solar.  Key points are as follows.  First, 
as expected, equatorial regions are on average warmer than the
high latitudes at pressures less than a few bars.  At low pressures,
longitudinal variation is comparable to the latitudinal
variation.  Second, the temperature declines smoothly with altitude
from $\sim3\,$bars to $\sim10\,$mbar.  This has important implications 
for spectra and light curves, which originate within this layer.
Third, dynamics has modified the deep stable radiative layer from 
10--100 bars, leading to significant latitude variation of both 
temperature and static stability, with warm poles and a cold equator.

Our HD 189733b simulations remain convectively stable everywhere throughout
the atmosphere.  This is illustrated in Fig.~\ref{hd189-spaghetti}, 
bottom panel, which shows the potential temperature $\theta$ versus 
pressure for the same profiles depicted in the top panel.  Potential 
temperature is a measure of entropy; an atmosphere that is neutrally 
stable to convection has $\theta$ constant with height, whereas 
$\theta$ increases with altitude in a statically
stable atmosphere \citep[e.g.][pp. 166-172]{salby-1996}.  As can be
seen in Fig.~\ref{hd189-spaghetti}, bottom panel, $\theta$ increases with
altitude in all the profiles.  This is true even 
on the nightside.  Nightside cooling (which tends to be stronger at low
pressure than high pressure) reduces the static stability of the nightside
profiles relative to the dayside profiles, 
but in our simulations this effect is insufficient to generate
a detached convection layer on the nightside.

\begin{figure}
\vskip 10pt
\includegraphics[scale=0.5, angle=0]{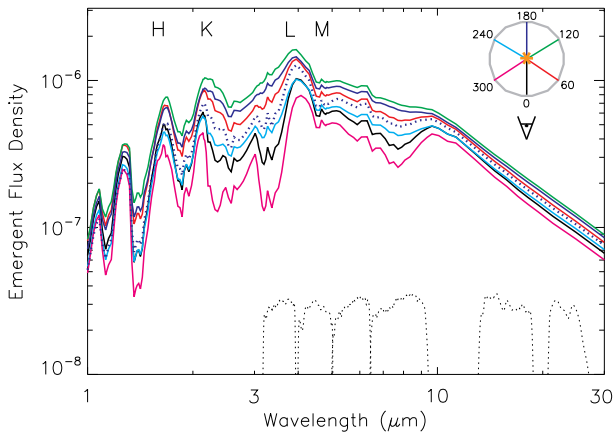}
\caption{Emergent flux density (${\rm ergs}\,\,\,{\rm s}^{-1}\, 
{\rm cm}^{-2}\,{\rm Hz}^{-1}$) from our nominal, solar-abundance
simulation of HD 189733b at six orbital phases.  {\it Black}, nightside,
as seen during transit; {\it red}, $60^{\circ}$ after transit;
{\it green}, $120^{\circ}$ after transit; {\it dark blue}, dayside,
as seen during secondary eclipse; {\it light blue}, $60^{\circ}$ after
secondary eclipse; and {\it magenta}, $120^{\circ}$ after secondary
eclipse. The key in the top right corner is color-coded with the spectra
to illustrate the sequence.   Thin dotted black lines at the bottom of
the figure show normalized {\it Spitzer} bandpasses and the letters at
the top show locations of the $H$, $K$, $L$, and $M$ bands.  This is
the same simulation as in Fig.~\ref{hd189-solar}.
For comparison, the dotted curve is a spectrum from a 1D planetwide
average radiative-equilibrium model.}
\label{hd189-spectra}
\end{figure}

Simulations of HD 189733b performed using opacities corresponding
to 5 and 10 times solar abundances exhibit similar behavior to
our solar case depicted in Figs.~\ref{hd189-solar}--\ref{hd189-spaghetti}.
The primary difference is that, compared to the solar case, the 5 and
10-times-solar cases have slightly warmer daysides, slightly cooler
nightsides, and smaller eastward offsets of the hot region from the
substellar point, especially at low pressure.  At 1 bar, in all three cases,
the approximate centroid of the hottest regions lie $\sim80^{\circ}$
east of the substellar point.  At 1 mbar, however,
the offset drops to $\sim50^{\circ}$ in the solar case but only
$\sim20^{\circ}$ in the 5 and 10-times-solar cases.  These differences
can be attributed to the greater opacities in the 5 and 10-times-solar
cases, which lead to greater heating rates and hence shorter radiative
time constants.  Despite the differences, we emphasize that all three cases
exhibit extremely similar {\it qualitative} temperature and wind patterns.

\subsection{Spectra and light curves}

We now turn to spectra and light curves.  Because our simulations
couple dynamics to radiative transfer, our model output includes
the emergent spectral flux versus wavelength everywhere on the
model grid at each timestep.  However, because we use a relatively
small number ($N_{\rm \lambda}=30$) of correlated-k wavelength bins, 
any spectra calculated from this output would have coarse spectral resolution.
Thus, we instead re-calculated the emergent fluxes offline 
at greater spectral resolution
($N_{\rm \lambda}=196$).  To do so, we took the 3D temperature output
at a given time and ran each vertical $T(p)$ column 
through our 1D spectral solver to calculate upward and downward fluxes
versus wavelength, taking care to use
the same opacity database and stellar model as used in the original 3D 
simulation.  The resulting emergent fluxes are essentially identical to
those generated by the 3D model except that they have higher spectral
resolution.  We then calculate disk-integrated spectra and light curves
using the procedures described in \citet{showman-etal-2008a} and
\citet{fortney-etal-2006b}.  Planetary limb darkening (or brightening), 
as viewed by the observer, is accounted for.

Figure~\ref{hd189-spectra} shows the resulting spectra for our solar-abundance
HD 189733b simulation at six orbital
phases.  Because the simulated temperatures decrease with
altitude in the relevant pressure range ($\sim0.01$--$1\,$bar), spectral
features are seen in absorption (rather than emission) throughout the orbit.  
Near the time of transit ({\it black curve}), the nightside faces 
Earth, with its cool temperatures and large vertical temperature 
gradients (Fig.~\ref{hd189-spaghetti}).  
This leads to relatively low fluxes with deep absorption bands of H$_2$O and
CH$_4$.  The hotter dayside ({\it solid dark blue curve}) 
has a smaller temperature gradient, leading to greater fluxes and
shallower absorption bands.  Interestingly, the fluxes $60^{\circ}$ of phase
before transit ({\it magenta}) are smaller than those 
occurring at transit, while the fluxes $60^{\circ}$ of phase before 
secondary eclipse ({\it green}) exceed those occurring immediately 
before/after the secondary eclipse itself.  This
phase offset results directly from the fact that the hottest and coldest
regions are shifted eastward from the substellar and antistellar points,
respectively (see Fig.~\ref{hd189-solar}).

\begin{figure}
\vskip 10pt
\includegraphics[scale=0.5, angle=0]{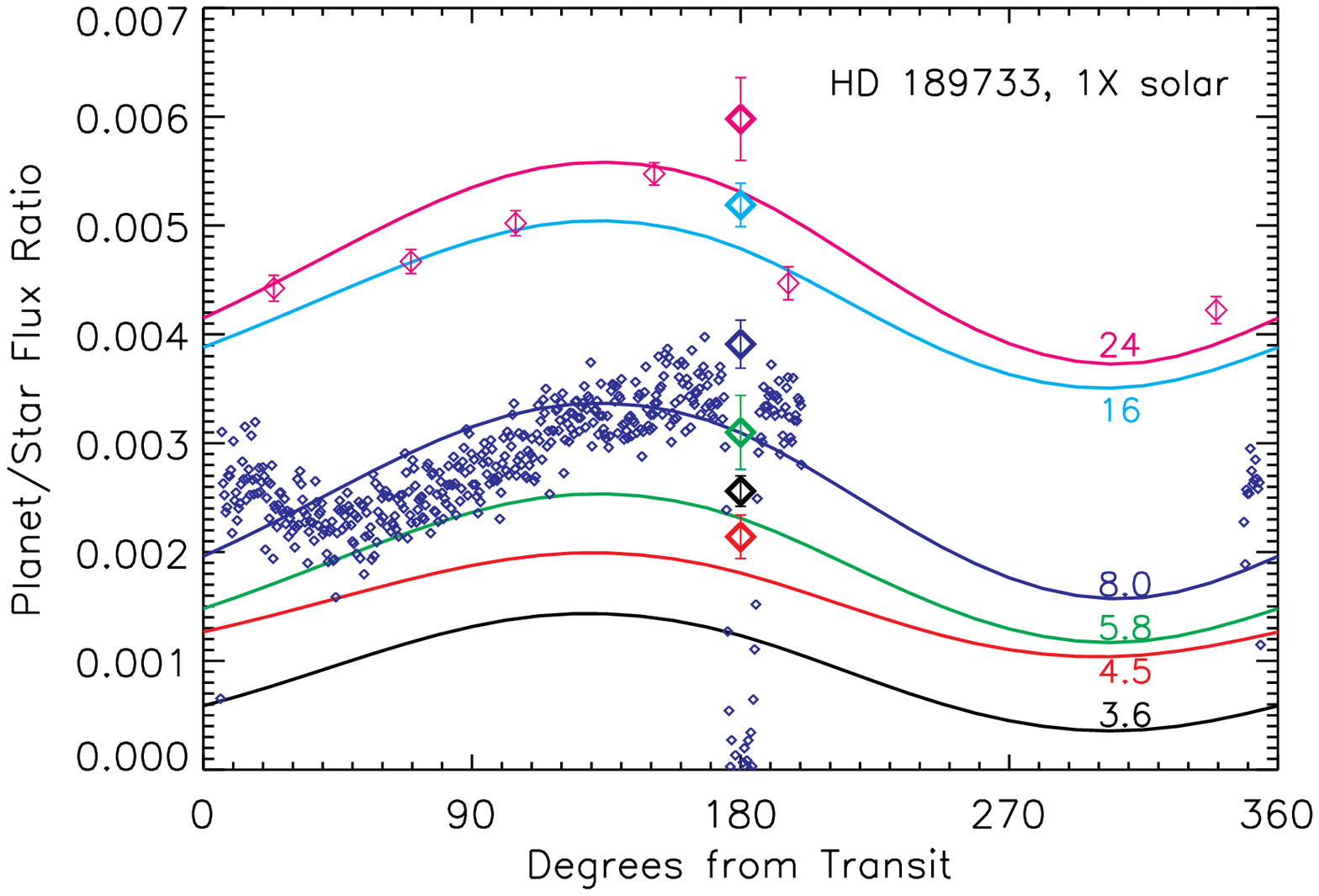}
\includegraphics[scale=0.5, angle=0]{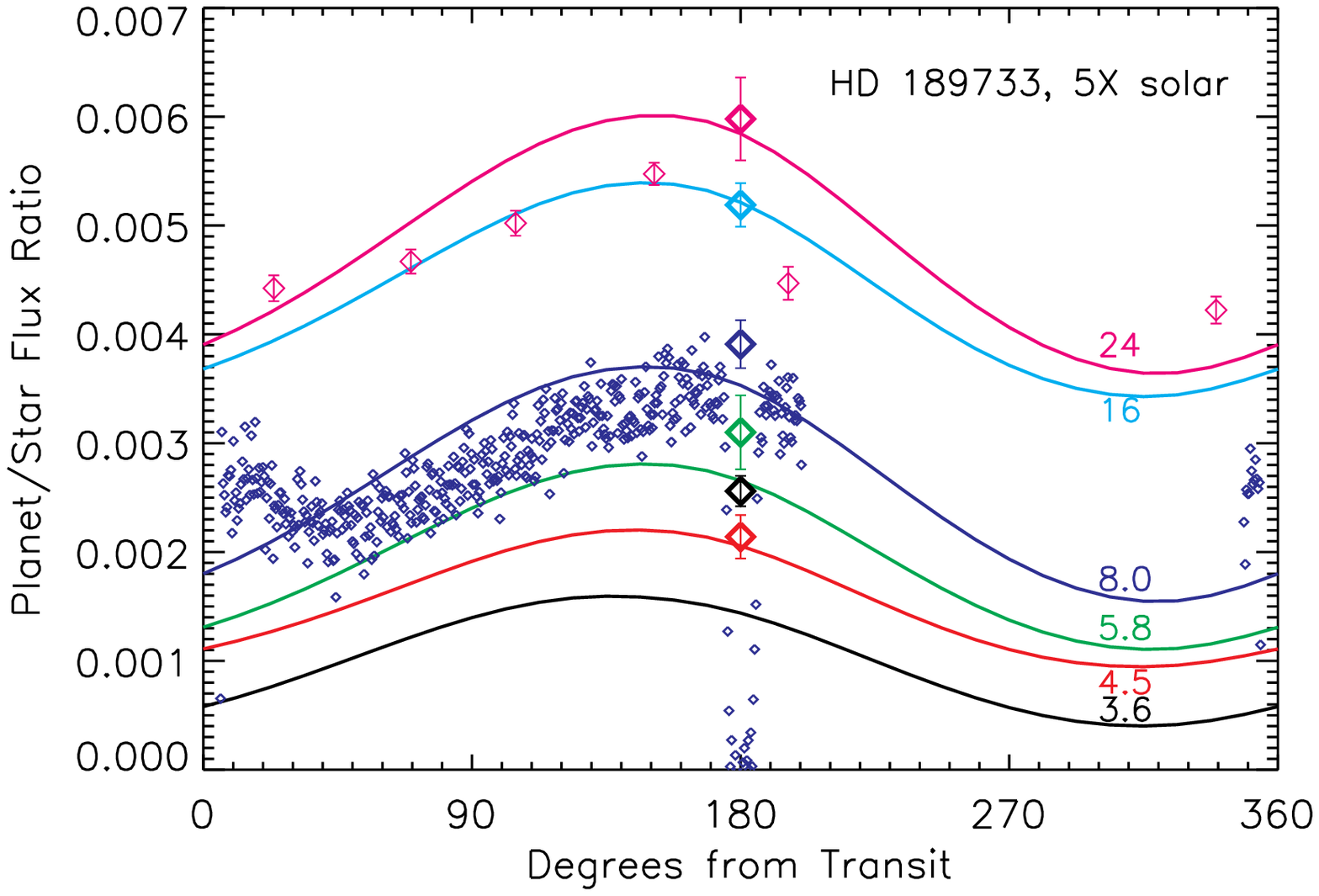}
\caption{Light curves versus orbital phase calculated in {\it Spitzer} 
bandpassesfor HD 189733b.  Top and bottom panels show light curves for 
our simulations using solar and five times solar abundances in the opacities, 
respectively.  Within each panel, moving from bottom
to top, the light curves are for wavelengths $3.6\,\mu$m ({\it black}),
$4.5\,\mu$m ({\it red}), $5.8\,\mu$m ({\it green}), 
$8\,\mu$m ({\it dark blue}), $16\,\mu$m ({\it light blue}), and 
$24\,\mu$m ({\it magenta}), respectively.  Overplotted is the 
{\it Spitzer} $8\,\mu$m light curve from \citet{knutson-etal-2007b}
in dark blue points and the binned $24\,\mu$m light curve from 
\citet{knutson-etal-2009a} in small magenta diamonds.  {\it Spitzer}
secondary-eclipse depths from \citet{charbonneau-etal-2008} and 
\citet{deming-etal-2006} are plotted at $180^{\circ}$ phase in large diamonds, 
with wavelengths color-coded as described above. Both simulations
have a resolution of C32 with 40 layers (top panel is same simulation
as in Fig.~\ref{hd189-solar}).}
\label{hd189-lightcurves}
\end{figure}

Figure~\ref{hd189-lightcurves} shows light curves calculated in 
{\it Spitzer} bandpasses for our HD 189733b simulations with opacities 
corresponding to solar ({\it top panel}) and five times solar 
({\it bottom panel}) abundances, respectively.  Black, red, green, dark blue, 
light blue, and magenta show the simulated light curves
at 3.6, 4.5, 5.8, 8, 16, and $24\,\mu$m.   Overlaid are the 
\citet{knutson-etal-2007b} 8-$\mu$m light curve in small blue dots 
and the binned \citet{knutson-etal-2009a}
24-$\mu$m light curve in small magenta diamonds, along with the 
{\it Spitzer} secondary-eclipse depths from \citet{charbonneau-etal-2008}
and \citet{deming-etal-2006} in large diamonds.

Overall, our simulated light curves (Fig.~\ref{hd189-lightcurves}) 
compare favorably to the observed ones. 
We are able to reproduce the modest day-night flux variation
seen in the observations at both 8 and $24\,\mu$m; 
this contrasts with our earlier simulations
using Newtonian heating/cooling \citep{showman-etal-2008a}, which
greatly overpredicted the day-night flux variation.  In our current
solar-opacity simulations, the ratio of maximum-to-minimum flux (within
a given {\it Spitzer} channel) ranges 
from 1.4 to 3.5 depending on wavelength, while in the five-times-solar
case it ranges from 1.6 to 4.1,
with the greatest flux ratios occurring at  $3.6\,\mu$m and the 
smallest at 16 and $24\,\mu$m.  Likewise, our simulated light curves
reach their peak flux before the secondary eclipse, a feature shared
by both the 8- and 24-$\mu$m light curves \citep{knutson-etal-2007b,
knutson-etal-2009a}. In the solar case ({\it top panel}), 
the offsets are close to $50^{\circ}$,
whereas at five times solar ({\it bottom panel}), the offsets range from
$\sim26^{\circ}$ at $24\,\mu$m to $\sim42^{\circ}$ at $3.6\,\mu$m.
 In the simulations, this phase offset results
directly from the eastward displacement of the
hottest and coldest regions from the substellar and antistellar points,
respectively (Fig.~\ref{hd189-solar}).  This phenomenon also occurred
in our previous simulations forced with Newtonian heating/cooling
\citep{showman-etal-2008a, cooper-showman-2005, showman-guillot-2002}.

We emphasize that the simulated light curves in Fig.~\ref{hd189-lightcurves}
are not fits to the observations; beyond choosing the metallicity,
no tuning of any kind was performed.  Instead, Fig.~\ref{hd189-lightcurves}
displays the natural interaction of radiation and dynamics as resolved by
the model.  Indeed, by explicitly representing both
the dynamics and the radiation, our goal here is to eliminate the 
tunable knobs that have been used to parameterize dynamics and/or 
radiation in some previous studies.

Nevertheless, there exist some important discrepancies between the simulated
and observed light curves. First, 
we do not reproduce the flux minimum that occurs 
$\sim50^{\circ}$ after transit in the observed 8-$\mu$m light curve.  
If real, this feature suggests the existence
of a local cold region to the {\it west} of the antistellar point
\citep{knutson-etal-2007b}.  However, this flux minimum is absent in the
{\it Spitzer} 24-$\mu$m light curve \citep{knutson-etal-2009a}, and
analysis of these data suggest instead that the minimum flux region
actually lies {\it east} of the antistellar point, which would be
qualitatively consistent with our simulations.  This 
hemisphere of the planet's surface is not well-resolved by the existing 
light curves, which cover only half an orbit; more data are needed to 
resolve the issue of where the true flux minima lie \citep{knutson-etal-2009a}.

Second, the phase offsets
in our simulations are somewhat too large, especially in our solar-opacity 
case.  The observed offsets of the flux peak are 
$16^{\circ}\pm6^{\circ}$ at $8\,\mu$m and 20--$30^{\circ}$
at $24\,\mu$m. In contrast, 
in the solar abundance simulation ({\it top panel}), the 
phase offsets are close to 
$50^{\circ}$ at all {\it Spitzer} bandpasses.   The agreement is better in
5-times-solar case, where the simulated offsets are $30^{\circ}$ at $8\,\mu$m
and only $26^{\circ}$ --- perfectly consistent with the observed offset --- 
at $24\,\mu$m.  On the other hand, at $24\,\mu$m, the solar case provides
an overall better match to the {\it magnitudes} of the light curve flux
values. 

\begin{figure}
\vskip 10pt
\includegraphics[scale=0.5, angle=0]{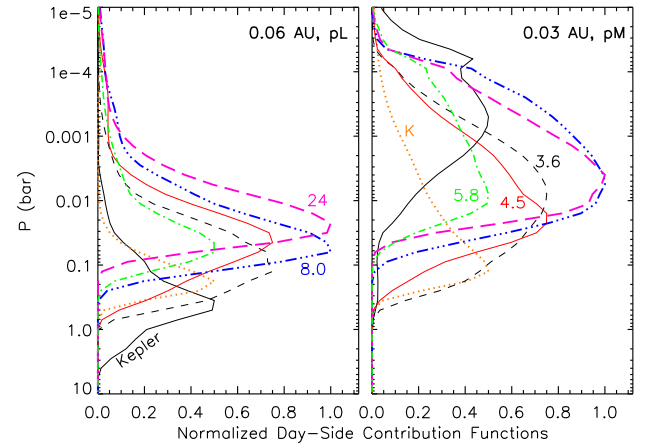}
\caption{Contribution functions 
\citep[e.g.][]{chamberlain-hunten-1987, knutson-etal-2009a}
calculated using our 1D radiative
transfer model for a generic cloud-free pL-class planet without atmospheric
TiO and VO ({\it left}) and a generic pM-class planet with atmospheric
TiO and VO ({\it right}).  Both are for dayside conditions, and 
both assume solar metallicity with equilibrium
chemistry.  Contribution
functions are calculated for various {\it Spitzer} broadband
filters (black short dashed, red solid, green dashed-dotted,
blue dashed-triple-dotted, and pink long-dashed curves
for 3.6, 4.5, 5.8, 8, and $24\,\mu$m, respectively), 
$K$ band (orange dotted curve), and the {\it Kepler} band at 450--900 nm
(black solid curve).  For clarity some of the curves have been normalized
to 0.5 or 0.75 rather than 1.}
\label{contribution-fcns}
\end{figure}

\begin{figure}
\vskip 10pt
\includegraphics[scale=0.5, angle=0]{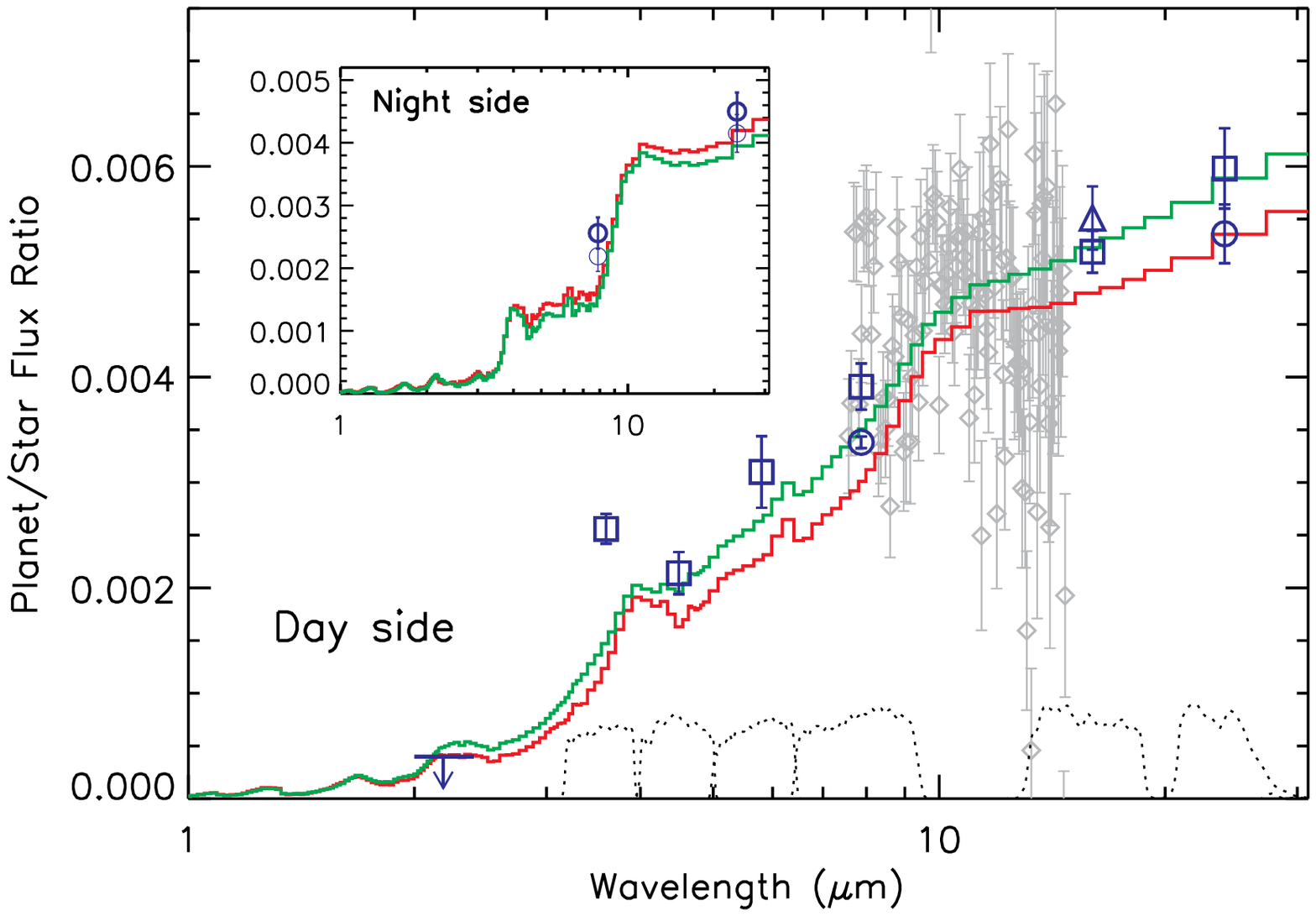}
\caption{Planet-to-star flux ratios for our HD 189733b 
simulations at the time immediately before/after secondary eclipse.  Red
and green depict our solar and 5 times solar cases, respectively.
Triangle shows the $16\,\mu$m IRS point from
\citet{deming-etal-2006}. Blue squares show data  
from \citet{charbonneau-etal-2008}, including their re-analysis of the
\citet{deming-etal-2006} data at $16\,\mu$m.  
Circles at 8 and $24\,\mu$m give secondary-eclipse depths obtained
from the light curves of \citet{knutson-etal-2007b, knutson-etal-2009a}.
Line at $2.2\,\mu$m ($K$ band) gives the 
upper limit from \citet{barnes-etal-2007}.
Grey points give the {\it Spitzer} IRS spectrum 
from \citet{grillmair-etal-2007}. {\it Inset:} Planet-to-star
flux ratios on the nightside for these same models.  Thin blue circles
show the nightside flux ratios at 8 and $24\,\mu$m as obtained from
the light curves of \citet{knutson-etal-2007b, knutson-etal-2009a}.
Thick blue circles are those same data corrected for the effect of
starspots.}
\label{hd189-planet-to-star}
\end{figure}

We now turn to the secondary-eclipse observations (large diamonds
in Fig.~\ref{hd189-lightcurves}).  We match reasonably well
the eclipse depths at 4.5, 5.8, 8, 16, and $24\,\mu$m, especially with our 
five-times-solar-metallicity model, although it would appear that 
our solar model is slightly too cool.  The greatest discrepancy in
both models is that we underpredict the 
3.6-$\mu$m secondary eclipse depth, by factors of 2.1 and 1.8 in our
solar and five-times-solar cases, respectively.  Our radiative-transfer
calculations suggest that $3.6\,\mu$m samples pressures of $\sim0.1$--1 bar, deeper than
any other {\it Spitzer} bandpass (see contribution functions 
in Fig.~\ref{contribution-fcns}, {\it left panel}).  
One possibility is that our model is too opaque at this bandpass;
a lower opacity would allow photons to escape from greater pressures, where
temperatures are hotter (Fig.~\ref{hd189-spaghetti}).   Fits
to the spectra of T dwarfs with the same atmospheric chemistry
and radiative transfer and similar effective temperatures also suggest
too much model opacity in this IRAC bandpass \citep{geballe-etal-2009}.
Alternatively, our model could simply be too cold in the 0.1--1-bar region, 
at least on the dayside.  Interestingly, published 1D models of HD 189733b
also have difficulty matching the $3.6\,\mu$m point \citep{barman-2008,
knutson-etal-2009a}.  A full light curve at $3.6\,\mu$m, possible
with warm {\it Spitzer}, would be invaluable in constraining this deep
layer of the atmosphere.

Figure~\ref{hd189-planet-to-star} reiterates these points by displaying
the planet-to-star flux ratio versus wavelength.  Our solar and five-times-solar
cases are depicted in red and green, respectively.  Photometric data
are shown in blue.  The reasonable
agreement at 4.5, 5.8, 8, 16, and $24\,\mu$m is evident, as is the discrepancy
at $3.6\,\mu$m.  Interestingly, this disrepancy is worse in our model
than in published 1D calculations.   On the other hand, unlike 
current 1D dayside models \citep[][]
{barman-2008, knutson-etal-2009a}, we are able to satisfy the
upper limit at $2.2\,\mu$m from \citet{barnes-etal-2007}.
On the nightside (see inset), we fare 
reasonably well against
the 8-$\mu$m and 24-$\mu$m planet-to-star flux ratios from the light curves
({\it blue circles}).

\begin{figure}
\vskip 10pt
\includegraphics[scale=0.5, angle=0]{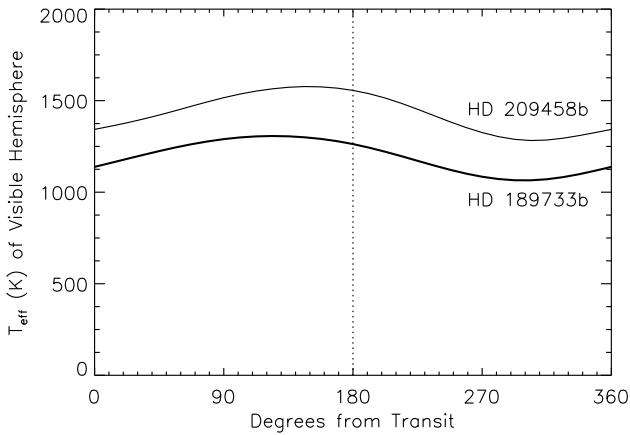}
\caption{Effective temperature versus orbital phase for the Earth-facing 
hemisphere of our nominal solar-opacity
simulations of HD 189733b (thick curve) and HD 209458b (thin curve).}
\label{Teff}
\end{figure}

Figure~\ref{Teff} shows the total luminosity versus orbital
phase for our solar-opacity simulation of HD 189733b expressed
as an effective temperature.  This was calculated by integrating
the planet's spectrum (Fig.~\ref{hd189-spectra}) over all wavelengths 
to obtain a flux, dividing by the Stefan-Boltzmann constant, and taking the
fourth root to obtain a temperature.  The effective temperature
reaches minima and maxima of $\sim1050$ and 1250 K, respectively,
and the phase offset is $\sim57^{\circ}$.  
As can be seen in Fig.~\ref{hd189-spectra}, much of this escaping radiation
lies in the 3--$5\,\mu$m range.

\begin{figure}
\vskip 10pt
\includegraphics[scale=0.5, angle=0]{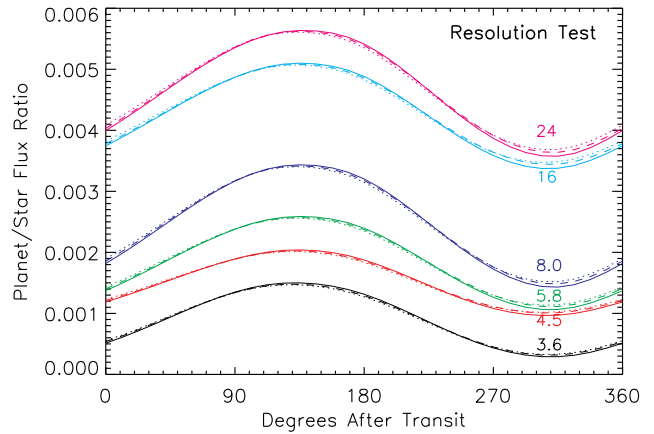}
\includegraphics[scale=0.5, angle=0]{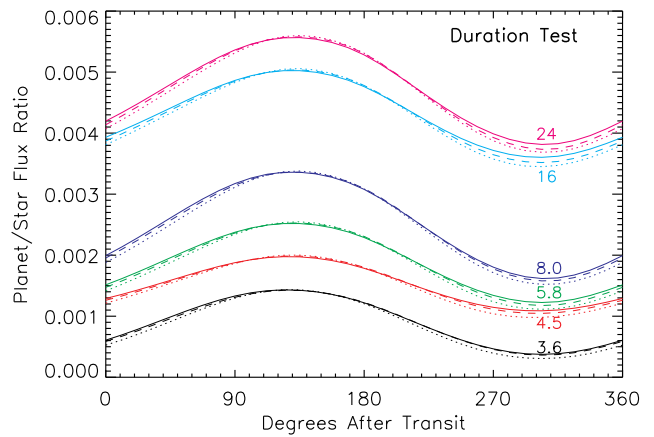}
\caption{Sensitivity of light curves, calculated in {\it Spitzer} bandpasses,
to model resolution ({\it top}) and integration time ({\it bottom}).  
Top panel shows HD 189733b simulations at horizontal resolutions of 
C64 (solid), C32 (dashed), and C16 (dotted), all at 301 Earth days of
integration time.  Bottom panel shows HD 189733b C16 simulations
at 104 ({\it dotted}), 995 ({\it dashed}), and 3968 ({\it solid})
Earth days.  All simulations adopt solar abudances
for the opacities.  Color scheme is as in Fig.~\ref{hd189-lightcurves}.}
\label{hd189-res-time-tests}
\end{figure}

Our basic results are insensitive to the model resolution and 
integration time, as illustrated in 
Fig.~\ref{hd189-res-time-tests}.  The top panel shows light
curves at {\it Spitzer} bandpasses for solar-metallicity HD 189733b simulations
performed at horizontal resolutions of C64 (solid), C32 (dashed), and C16 
(dotted) (approximately equivalent to global resolutions of $256\times128$,
$128\times64$, and $64\times32$ in longitude and latitude, respectively).
All are at an integration time of 301 Earth days and have 40 vertical layers.  
In the bottom panel are light curves from the low-horizontal-resolution (C16) 
case at integration
times of 104, 995, and 3968 Earth days.  The simulated light curves
are very similar in all these cases.  This indicates that our model
resolutions and integration times are sufficient to capture the dynamics.

\begin{figure}
\vskip 10pt
\includegraphics[scale=0.5, angle=0]{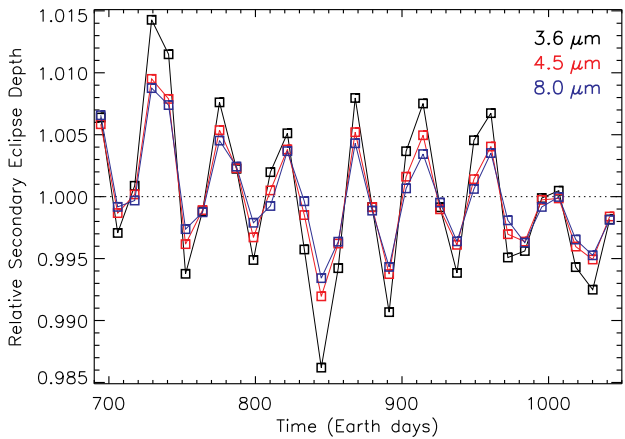}

\caption{Secondary-eclipse depths versus time calculated from
our synchronously rotating HD 189733b simulation with solar opacity.  
Curves show results at
the 3.6, 4.5, and 8-$\mu$m {\it Spitzer} IRAC bands in black, red,
and blue, respectively, each normalized to its mean.
Our predicted variability is $\sim1$--2\%, consistent
with the observational 8-$\mu$m upper limit from \citet{agol-etal-2008}.}
\label{hd189-variability}
\end{figure}

Our simulations are consistent with recent upper limits on the 
temporal variability of HD 189733b reported by \citet{agol-etal-2008}.
They analyzed five {\it Spitzer} IRAC 8-$\mu$m secondary-eclipse 
observations scattered over $\sim370$ Earth days and found 
that the variability in secondary-eclipse depth is less than 10\%
around their best-fit mean value of 0.00347.  To compare with this
observation, Fig.~\ref{hd189-variability} shows secondary-eclipse depths 
in the IRAC 3.6, 4.5, and 8-$\mu$m bands versus time, calculated from
our HD 189733b simulation (as in Figs.~\ref{hd189-lightcurves} and \ref{hd189-res-time-tests}) by properly integrating the simulated infrared spectra
over the {\it Spitzer} bandpasses.  Over a period of several hundred
days, our simulated variability is $\sim1\%$
at 4.5 and $8\,\mu$m and $\sim1.5\%$ at $3.6\,\mu$m, fully consistent
with the \citet{agol-etal-2008} upper limit.  Interestingly, much
of the predicted variability involves a coherent oscillation with a period of 
approximately 43 days, although a weak downward trend (corresponding
to a mean decrease in eclipse depth of about 0.5\%) is also present.
The predominance of a single oscillation period suggests the presence
of a global sloshing mode with a period of weeks; future work will be 
required to investigate this phenomenon and the extent to which it
may depend on parameters such as the atmospheric vertical structure.

\section{Non-synchronously rotating HD 189733b}

Several authors have suggested that hot Jupiters in near-circular
orbits will synchronously rotate because their expected spindown 
times are $\sim10^6\,$years for a Jupiter-like tidal $Q$ of $10^5$ 
\citep{guillot-etal-1996, lubow-etal-1997}.  While this is plausible, 
no observations currently exist to constrain the rotation period 
of any hot Jupiter.  Moreover, \citet{showman-guillot-2002} argued 
that dynamical
torques between the atmosphere and interior could lead to equilibrium
rotation rates that deviate modestly from synchronous.
Therefore, we ran several simulations 
of HD 189733b to investigate the role that non-synchronous rotation
has on the dynamics and resulting observables.

\begin{figure}
\vskip 10pt

\includegraphics[scale=0.6, angle=0]{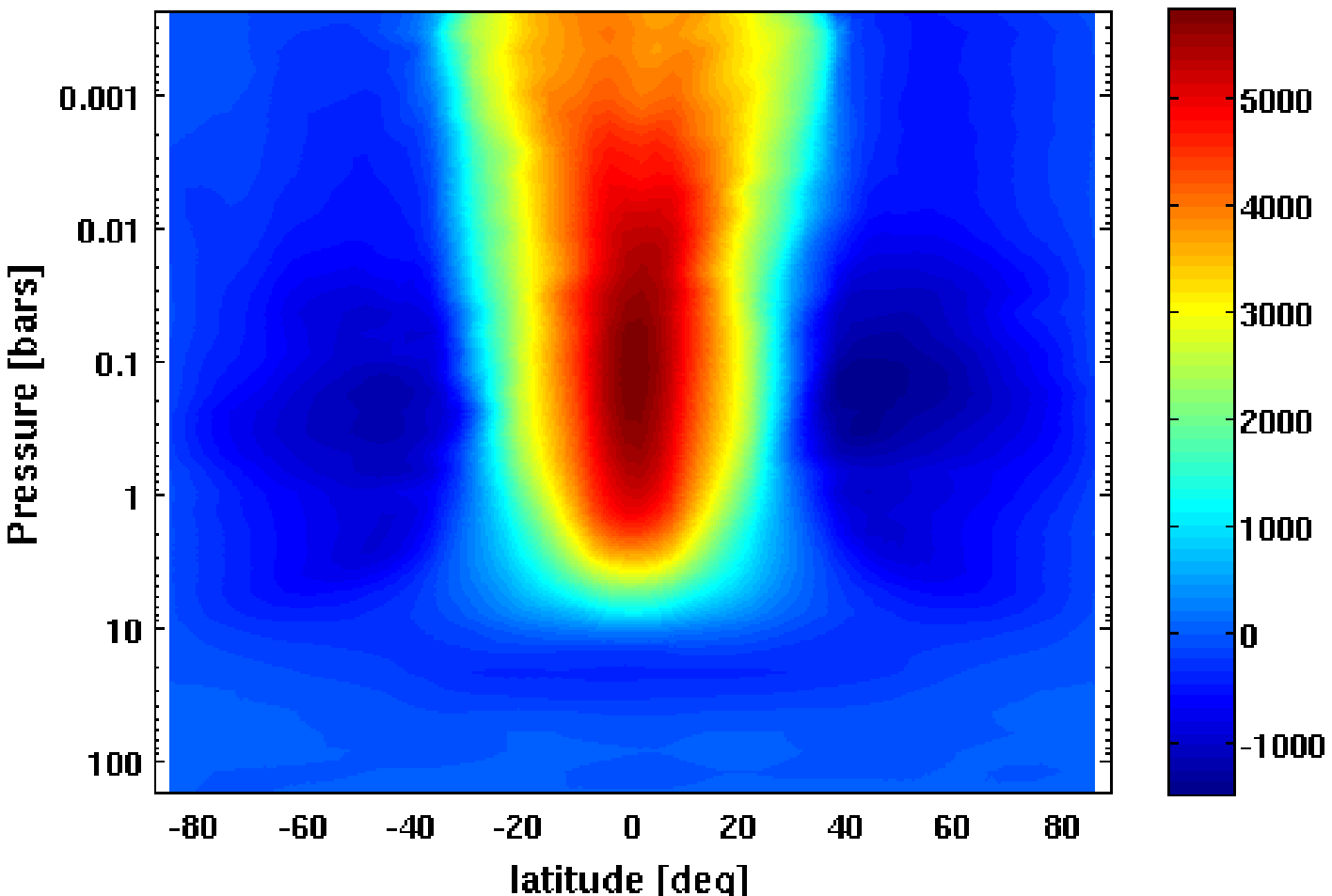}
\includegraphics[scale=0.6, angle=0]{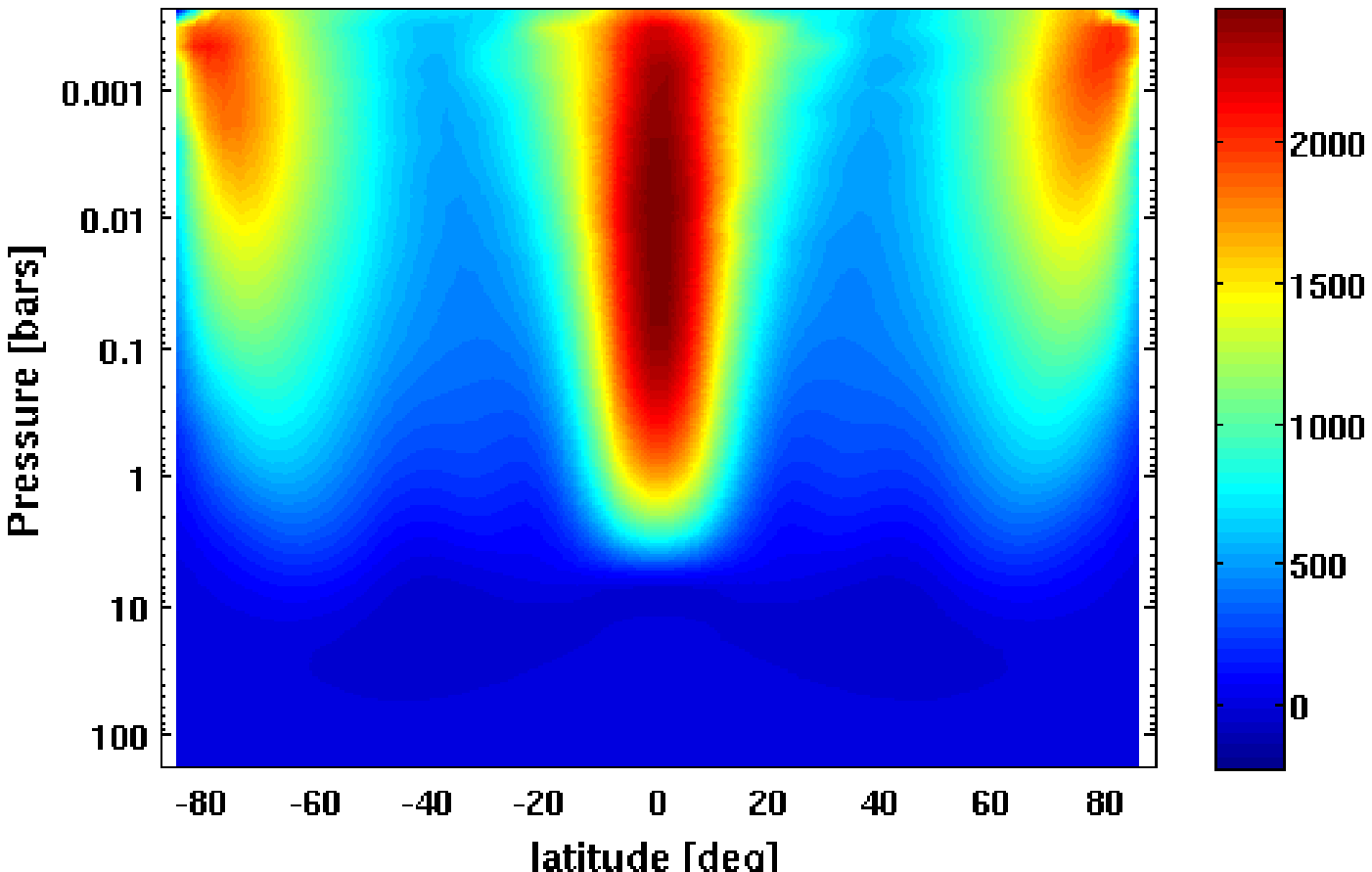}
\includegraphics[scale=0.6, angle=0]{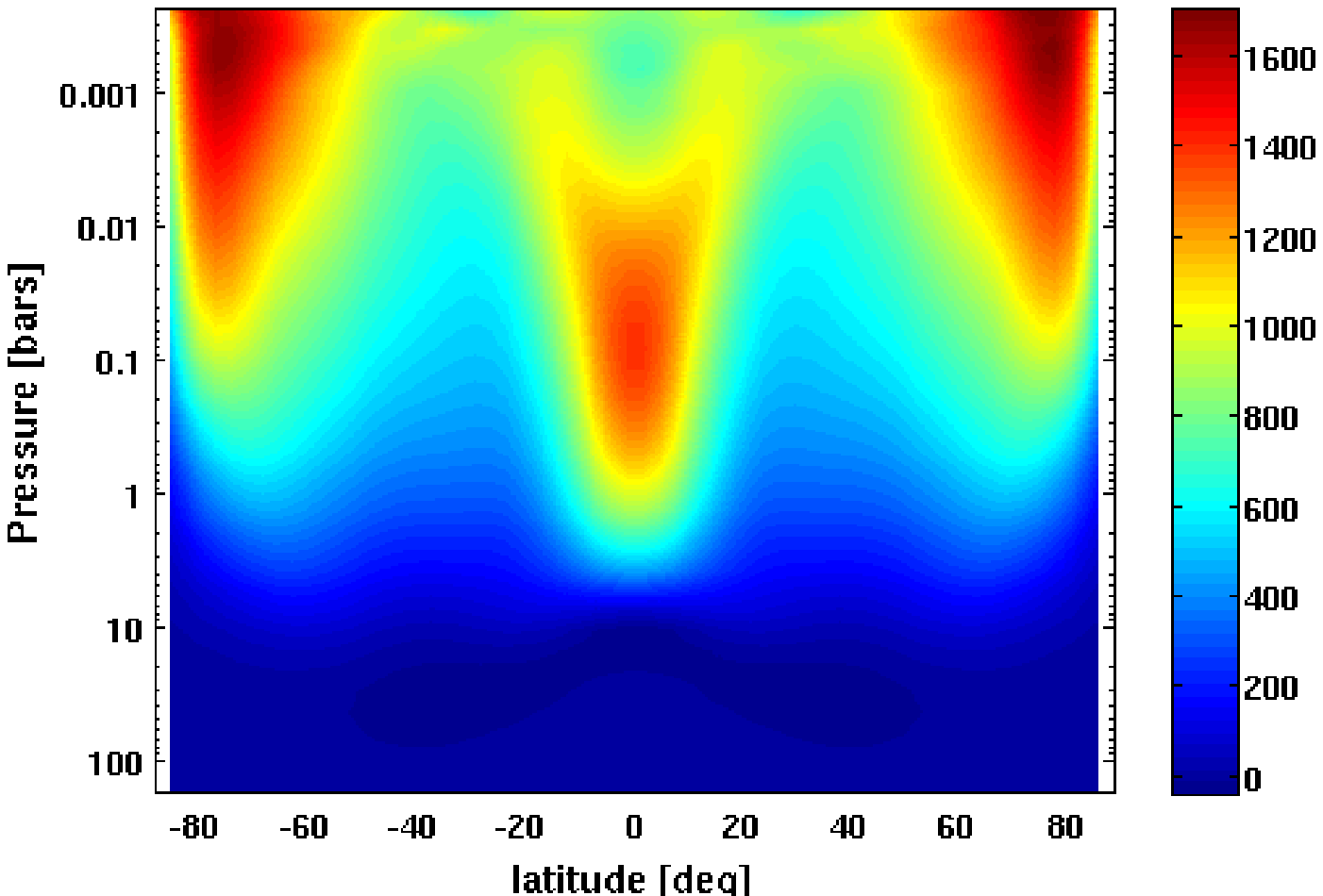}
\caption{Effect of non-synchronous rotation on jet structure.  Shows
zonal-mean zonal wind versus latitude and pressure for HD 189733b
cases with half ({\it top}), 1.5 times ({\it middle}), and twice
({\it bottom}) the synchronous rotation rate.  Opacities use solar
abundances.  Scalebar gives speeds in m$\,$s$^{-1}$. 
All simulations have a horizontal resolution of C32
with 40 layers and can be directly compared with Fig.~\ref{hd189-ubar}.}
\label{hd189-nonsynch-ubar}
\end{figure}

Figure~\ref{hd189-nonsynch-ubar} shows the zonal-mean zonal
winds versus latitude and pressure for cases with rotation rates
$\Omega$ that are half, 1.5, and twice the synchronous values
({\it top, middle, and bottom panels}, respectively). These
cases all assume solar metallicity and can be directly compared
to their synchronously rotating counterpart in Fig.~\ref{hd189-ubar}. 
As can be seen, the rotation rate has a significant effect on the
mean jet structure.  The slowly rotating case ({\it top}) qualitatively 
resembles the synchronous case, with an eastward equatorial jet and
westward high-latitude flows, except that the equatorial jet is narrower 
and faster.  At rotation rates faster than synchronous, however,
the equatorial jet weakens and strong eastward {\it polar} jets
develop --- a phenomena unseen in our synchronous cases.  At
rotation rates that are double the synchonous value ({\it bottom}),
the mean speed of the polar jets exceeds that of the equatorial jet.
Interestingly, these faster rotation rates are also accompanied
by a general slowdown in the wind speeds; at double the synchronous
rotation rate, the peak zonal-mean zonal wind speed is $\sim1.6\,{\rm km}
\,{\rm s}^{-1}$, half the value in the sychronous case.

Two effects can influence the jet structure in these
nonsynchronous cases.  First, in the nonsynchronous cases, the 
illumination sweeps in longitude as seen in the planet's rotating 
reference frame --- in contrast to the synchronous case, where
the illuminated region is locked to a specific range of longitudes.
In all cases (synchronous or not) the jets are presumably
accelerated by large-scale eddies generated by the
day-night heating gradient.  This must particularly be true
of the superrotating equatorial jet, which corresponds to a 
local maximum of angular momentum per mass and can only be
formed by up-gradient momentum transport by eddies.  Sweeping the
heating pattern in longitude can potentially change the character
of these jet-driving eddies and hence alter the resulting jet
structure.  This may explain, for example, the trend of decreasing
jet speed with increasing rotation rate (Fig.~\ref{hd189-nonsynch-ubar}).
Interestingly, the peak equatorial jet speed in an inertial
reference frame --- namely, the sum of the peak jet speed in 
Fig.~\ref{hd189-nonsynch-ubar} and the planetary rotation rate of
$\Omega a$ at the equator --- is nearly constant (to within $\sim15\%$)
in all the cases shown in Figs.~\ref{hd189-ubar} and \ref{hd189-nonsynch-ubar}.

Second, changing the rotation rate changes the strength of
the Coriolis force and, particularly, the strength of the
so-called ``$\beta$ effect,'' where $\beta= 2\Omega\cos\phi/a$ 
is the northward gradient of the Coriolis parameter.  For flows
driven primarily by creation of turbulence at small scales, 
turbulence theory predicts that jet widths in latitude scale as 
$\sim\pi(U/\beta)^{1/2}$, where $U$ is a mean wind speed 
\citep{rhines-1975, vasavada-showman-2005}.  This ``Rhines'' 
scaling suggests that slower (faster) rotation rates would lead to 
wider (narrower) jets.  In simulations of {\it synchronously rotating} 
hot Jupiters, \citet{showman-etal-2008a} showed that faster rotation 
indeed leads to narrower jets and vice versa --- but the dependence of jet
width on rotation rate is weaker than the
inverse-square-root Rhines prediction.  Presumably, this deviation 
occurs because, in the hot Jupiter case,
the jets are driven not at small scales but by eddies induced by
the day-night heating gradient, which have length scales comparable to the
jet widths.   The differences between our nonsychronous cases
(Fig.~\ref{hd189-nonsynch-ubar}) and the synchronous $\Omega$
parameter variations in \citet{showman-etal-2008a} suggest that
the trends in Fig.~\ref{hd189-nonsynch-ubar} cannot be explained 
solely by the $\beta$ effect but also depend on changes in the eddy 
behavior as discussed above.  We defer a detailed diagnosis of the dynamics
of this process to a future study.

\begin{figure}
\vskip 10pt
\includegraphics[scale=0.5, angle=0]{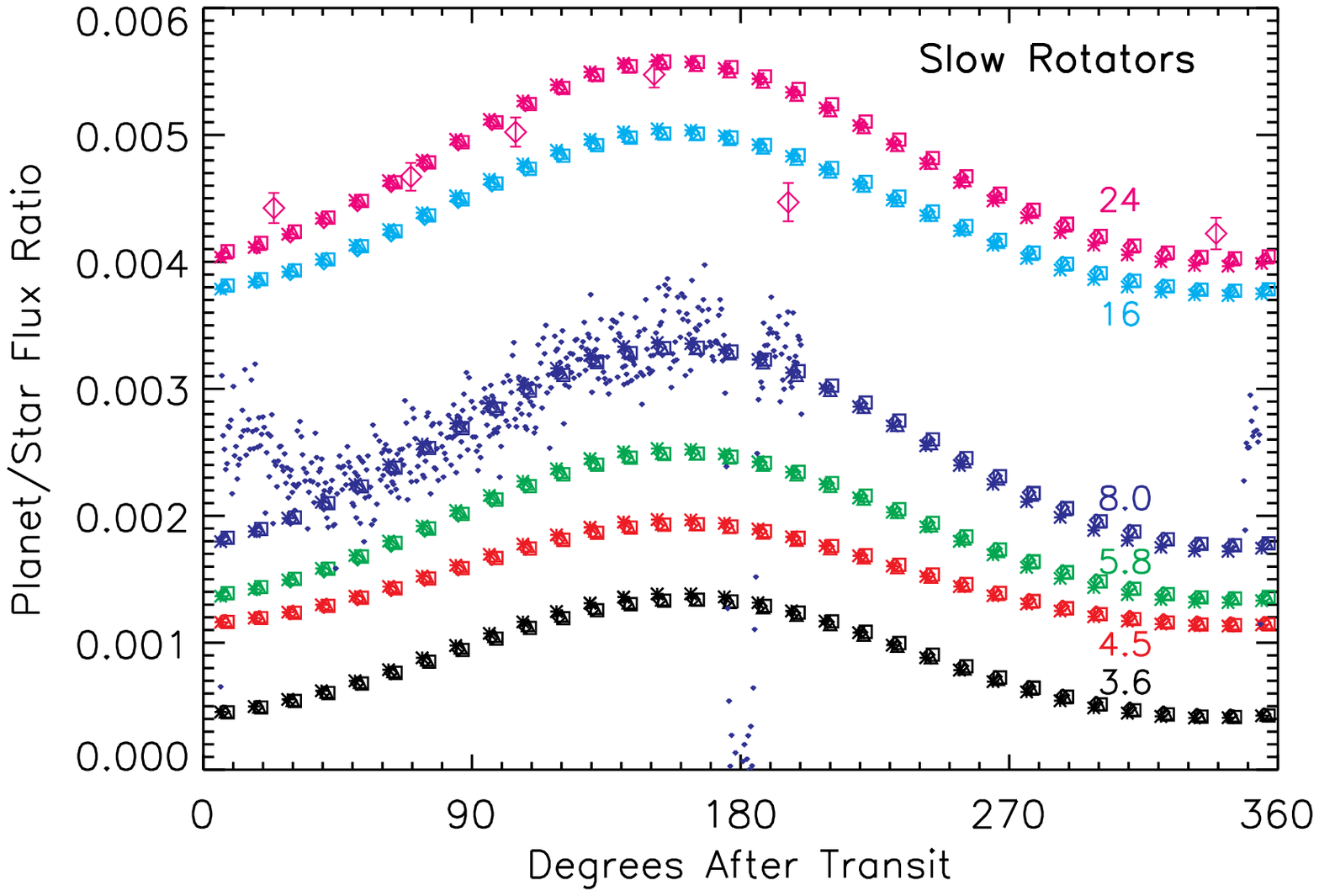}
\includegraphics[scale=0.5, angle=0]{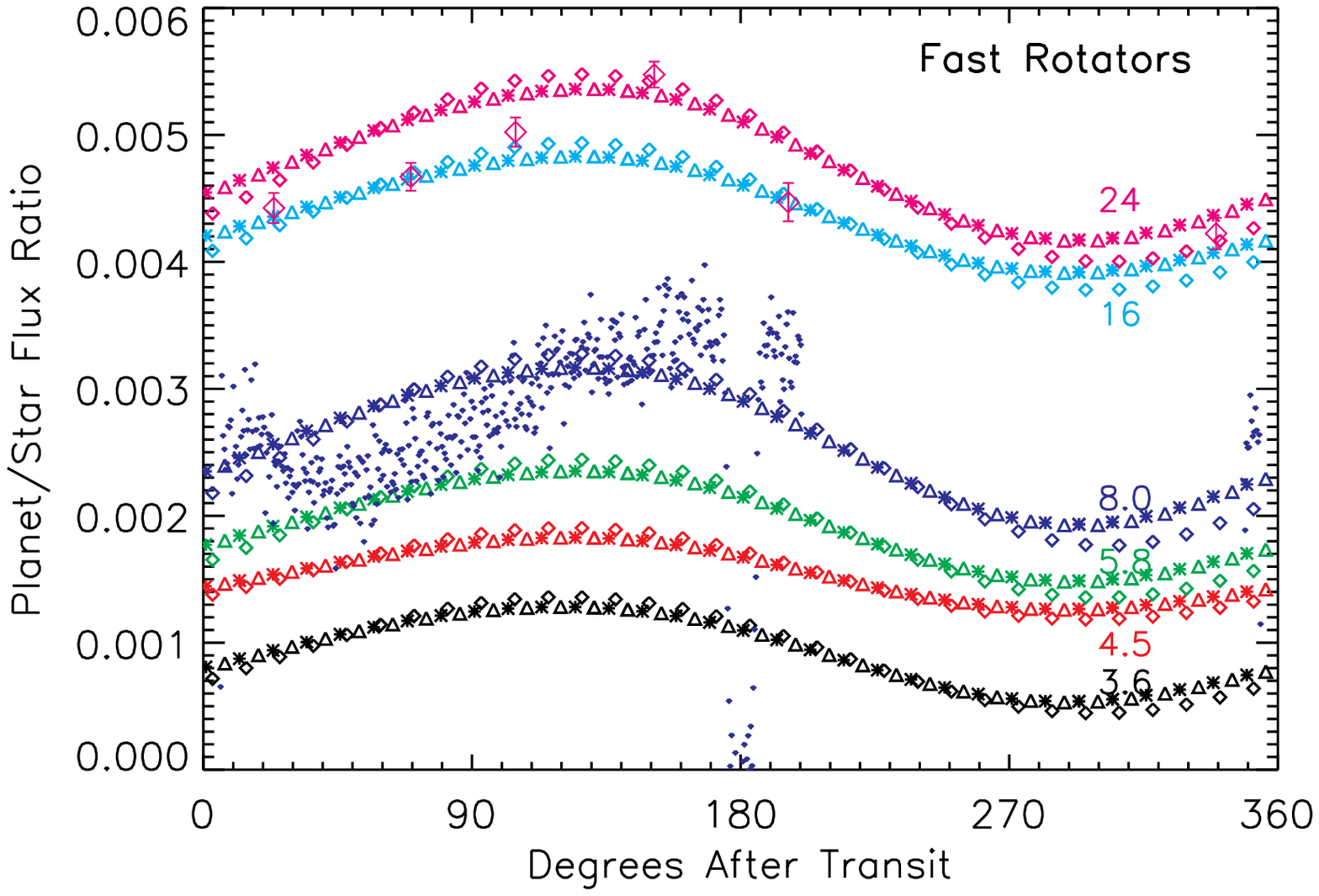}
\caption{Light curves for our non-sychronous, solar-metallicity 
HD 189733b simulations shown in Fig.~\ref{hd189-nonsynch-ubar}.  
Top panel shows case with half the synchronous rotation rate ({\it squares},
{\it asterisks,} and {\it triangles} denote different times).  
Bottom panel shows cases with 1.5 times the sychronous rotation rate
({\it triangles} and {\it asterisks} denoting different times) 
and twice the synchronous rotation rate ({\it diamonds}).}
\label{hd189-nonsynch-lightcurves}
\end{figure}

Figure~\ref{hd189-nonsynch-lightcurves} demonstrates that
nonsynchronous rotation has a significant effect on the light curves. 
Interestingly, the phase offsets are smaller for the slowly rotating case
($\Omega$ half the synchronous value; {\it top panel}) and larger
for the rapidly rotating cases ($\Omega$ 1.5 and two times the 
synchronous value; {\it bottom panel}).  This behavior makes sense
physically.  
As viewed in the synchronously rotating reference frame, the nonsynchronous
rotation is equivalent to a westward motion of the planetary interior
in slowly rotating case but an eastward motion of the planetary
interior in the rapidly rotating cases.  This rotation combines with 
the winds shown in Fig.~\ref{hd189-nonsynch-ubar}\footnote{The winds
shown in Fig.~\ref{hd189-nonsynch-ubar} are those in the 
planet's {\it nonsynchronous} reference frame;
to represent the winds in the {\it synchronously} rotating reference
frame, one must add $(\Omega - \Omega_{\rm syn})a\cos\phi$, where
$\Omega$ is the actual rotation rate,
$\Omega_{\rm syn}$ is the rotation rate of the synchronous reference
frame, $a$ is planetary radius, and $\phi$ is latitude.}
 to produce (as seen in the synchronous frame) a strong, 
latitude-averaged eastward flow in the rapidly rotating cases; in 
contrast, in the slowly rotating cases, the midlatitudes flow strongly 
westward, although the equatorial jet is still eastward.  
The upshot is that the hottest regions are shifted eastward
from the substellar point by a greater distance in the rapidly
rotating cases than in the slowly rotating case.  
This leads to a small phase
shift in the slowly rotating case and a large phase shift in the
rapidly rotating cases (Fig.~\ref{hd189-nonsynch-lightcurves}).

Interestingly, the {\it shape} of the lightcurve is also affected ---
the flux peak is narrower in the slowly rotating case than
in the rapidly rotating case.  This seems to occur because the
dayside hot regions span a smaller range of longitudes 
in the slowly rotating case than in the rapidly rotating cases.

The theoretical 8-$\mu$m light curve in the slowly rotating case 
actually provides quite a good match  to the observed 8-$\mu$m light 
curve, reproducing both the shape and phase offset of the flux peak.  
Nevertheless, we still do not reproduce the flux minimum seen after 
transit.  The rapidly rotating cases provide a significantly
poorer match at $8\,\mu$m, with a flux peak that is too broad and
a phase shift that is too large.  The situation is more ambiguous
at $24\,\mu$m; both cases provide reasonable though not perfect
matches to the observed light curve. We re-emphasize that these 
are not fits to data but rather are theoretical predictions.  
Moreover, although it could be tempting to infer
a planetary rotation rate from these comparisons,
this is premature, as other factors not explored here (e.g.,
disequilibrium chemistry) could also have large effects on
the light curves.

\section{Synchronously rotating HD 209458b}

\subsection{Circulation regime, spectra, and light curves}

{\it Spitzer} secondary-eclipse photometry shows that
the 4.5 and 5.8-$\mu$m brightness temperatures of HD 209458b reach
$1700$--$1900\,$K \citep{knutson-etal-2008a}.  These values significantly 
exceed both the planet's effective temperature and the 
inferred brightness temperatures at the 3.6, 8, and 24-$\mu$m bands, 
which are closer to $\sim1500\,$K.   Given these facts, 
\citet{knutson-etal-2008a} and \citet{burrows-etal-2007b, burrows-etal-2008}
suggested that the atmosphere of HD 209458b contains a thermal inversion 
(hot stratosphere) on the dayside, with temperatures reaching $\sim2000\,$K
or more.

Long before these {\it Spitzer} data were available, \citet{hubeny-etal-2003} 
and \citet{fortney-etal-2006a} showed that such inversions result
naturally from gaseous TiO and VO, which are extremely opaque in the visible
wavelength range and cause absorption of starlight at substantially
lower pressures ($\sim$mbar) than would occur in the absence of
these species.   Undetermined photochemical products are another possible
absorber \citep{burrows-etal-2007b, burrows-etal-2008}. 
\citet{zahnle-etal-2009}, for example, have suggested that disequilibrium
sulfur species produced from photochemical destruction of H$_2$S could provide
sufficient short-wavelength absorption to account for the stratospheres.

However, to date, radiative-equilibrium models of HD 209458b have been 
unable to reproduce the observed {\it Spitzer} photometry 
\citep{fortney-etal-2008}: while they successfully reproduce the 
high 4.5 and 5.8-$\mu$m fluxes, they overpredict the 3.6, 8, and 
$24\,\mu$m fluxes.  This suggests that the stratosphere predicted
in these radiative-equilibrium models is too hot and/or too broad in
vertical extent.  \citet{burrows-etal-2007b} 
obtained better agreement, especially at $3.6\,\mu$m, by including 
an {\it ad hoc} heat sink to mimic the effect of day-to-night heat transport 
by the atmospheric circulation.  By confining this sink to pressures 
between 0.01 and 0.1 bars, \citet{burrows-etal-2007b} were able to reproduce 
the low 3.6-$\mu$m flux while keeping the 4.5 and 5.8-$\mu$m fluxes high.
However, as \citet{showman-etal-2008a} pointed out, day-night heat transport is
unlikely to be confined to a narrow pressure range; in 
\citet{showman-etal-2008a}'s 3D circulation models, for example,
the dayside ``heat sink'' (expressed in ${\rm K}\,{\rm s}^{-1}$)
caused by dynamics increases monotonically with decreasing pressure
from $\sim10$ bars to the top of their model at $0.001\,$bar (their Fig.~10).  
Thus, an important question is whether a 3D dynamical model, 
coupled to realistic radiative transfer and including TiO/VO opacity, can match
the observed secondary-eclipse spectrum of HD 209458b.

Light curves are also of interest.  From several brief {\it Spitzer}
observations at different phases, \citet{cowan-etal-2007} placed a 
2-$\sigma$ upper limit of 0.0015 on the peak-to-peak phase variation
in the planet/star flux ratio at $8\,\mu$m.  
This suggests that the difference in the planet's day
and night 8-$\mu$m brightness temperatures is less than a few hundred K.  
In contrast, 
\citet{fortney-etal-2008} suggested that, as a pM-class planet,
HD 209458b should exhibit much greater phase variations 
in the mid-infrared than planets lacking 
atmospheric TiO/VO.  Physically, the idea is that the presence of TiO and VO
moves the photospheres upward to low pressure, where the radiative
time constants are short and the air can experience rapid dayside
heating and nightside cooling.  In contrast, planets without atmospheric
TiO/VO would have deeper photospheres, where the radiative time constants
are longer than typical advection times, thus yielding more modest day-night
phase variations.  \citet{fortney-etal-2008} also suggested that
phase offsets of the hottest/coldest regions from
the substellar/antistellar points would be smaller for pM-class planets
than for pL-class planets.  We here wish to test these ideas by determining
the circulation patterns and lightcurves for a model of HD 209458b
with TiO/VO opacity.  Note that even if hot-Jupiter stratospheres turn out
to result from shortwave absorption by compounds other than TiO and VO 
\citep[e.g.,][]{zahnle-etal-2009}, our simulations
will give a qualitative picture of how dynamics
responds to the presence of a compound that absorbs strongly in
the visible.

Thus, we performed several simulations of HD 209458b including TiO
and VO opacity in chemical equilibrium.  When temperatures are too
cold for TiO and VO to exist in the gas phase, they are absent, but
when temperatures are warm enough, they are included.  
 
\begin{figure}
\vskip 10pt
\includegraphics[scale=0.43, angle=0]{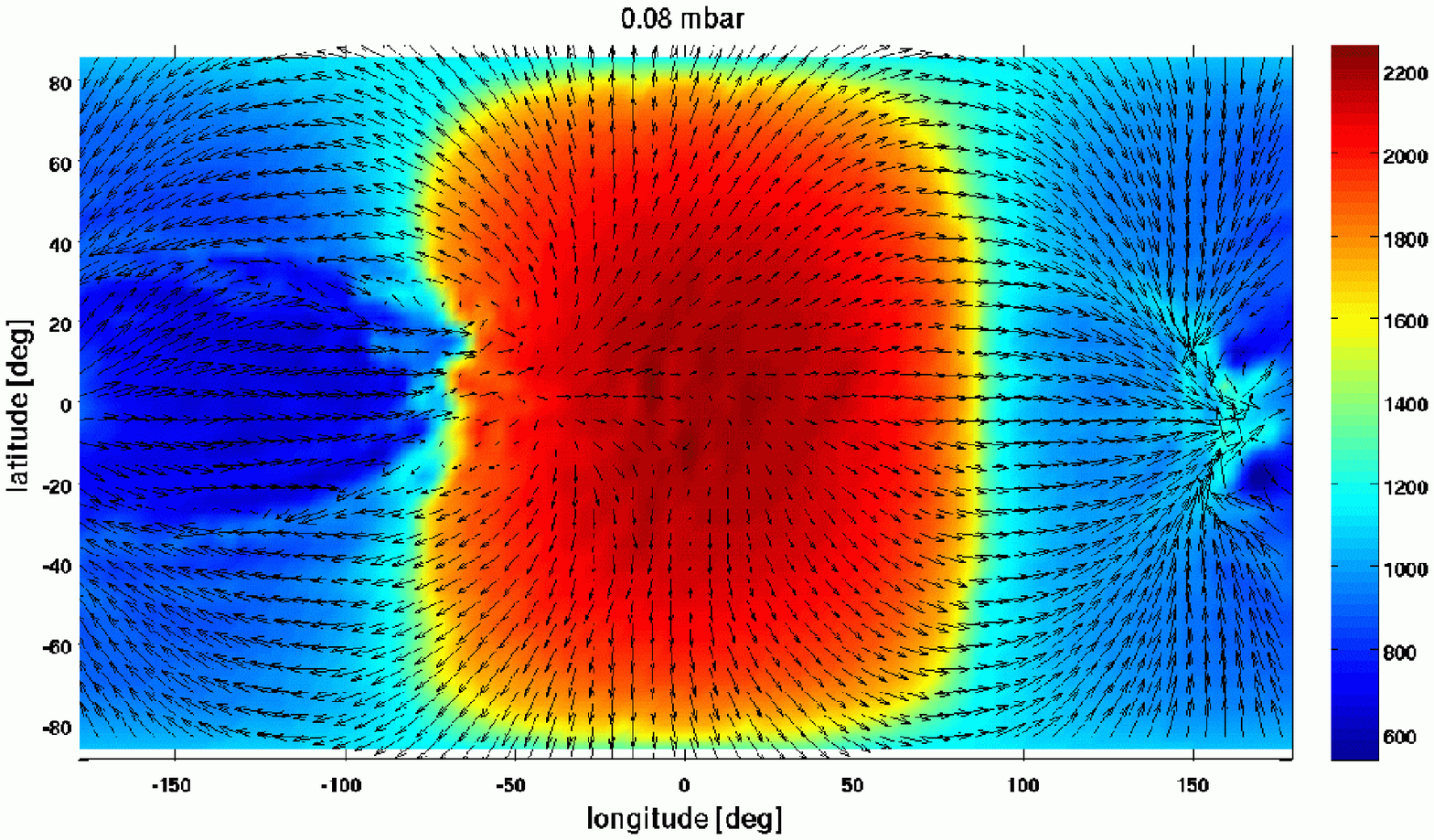}
\includegraphics[scale=0.43, angle=0]{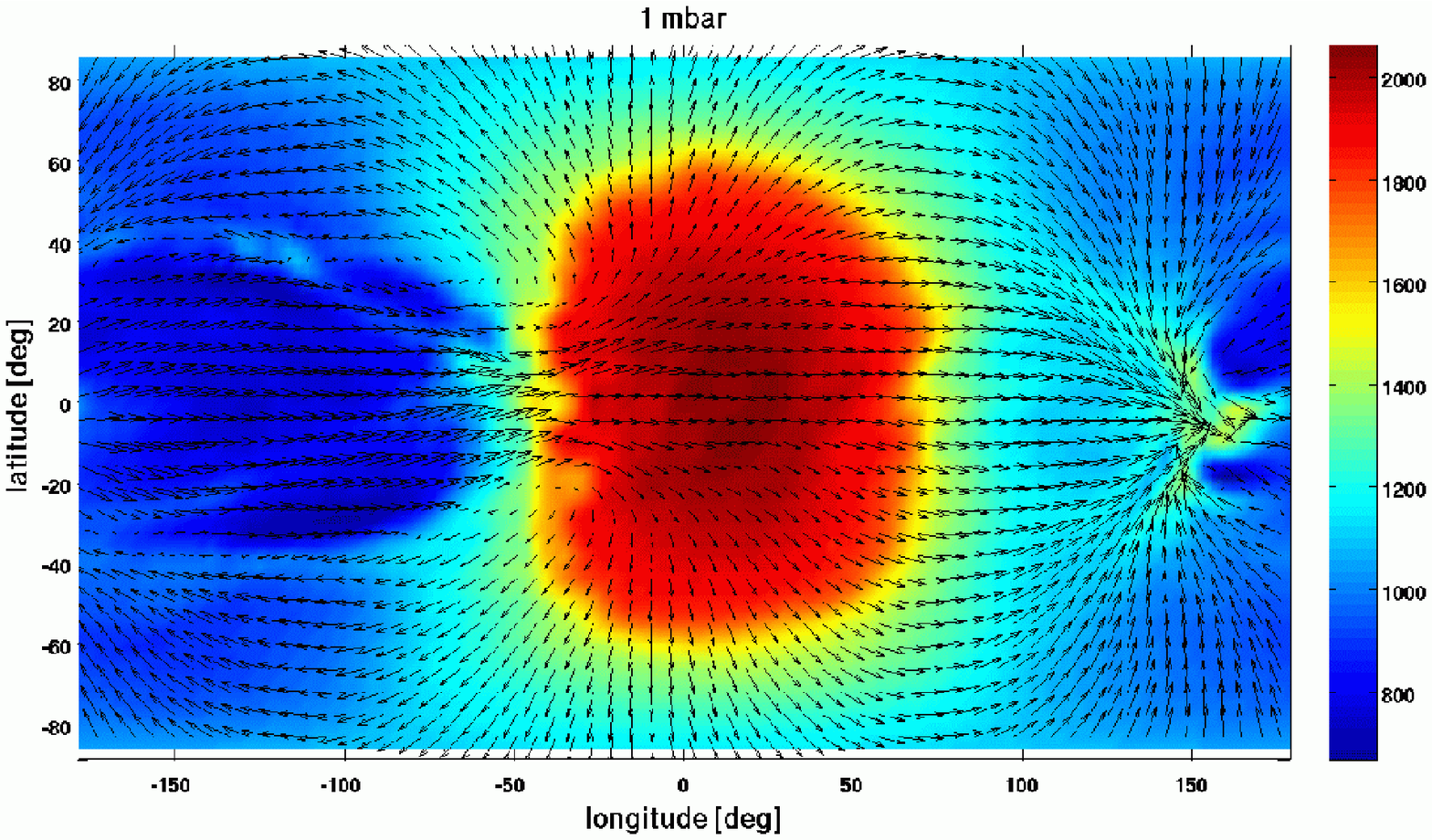}
\includegraphics[scale=0.43, angle=0]{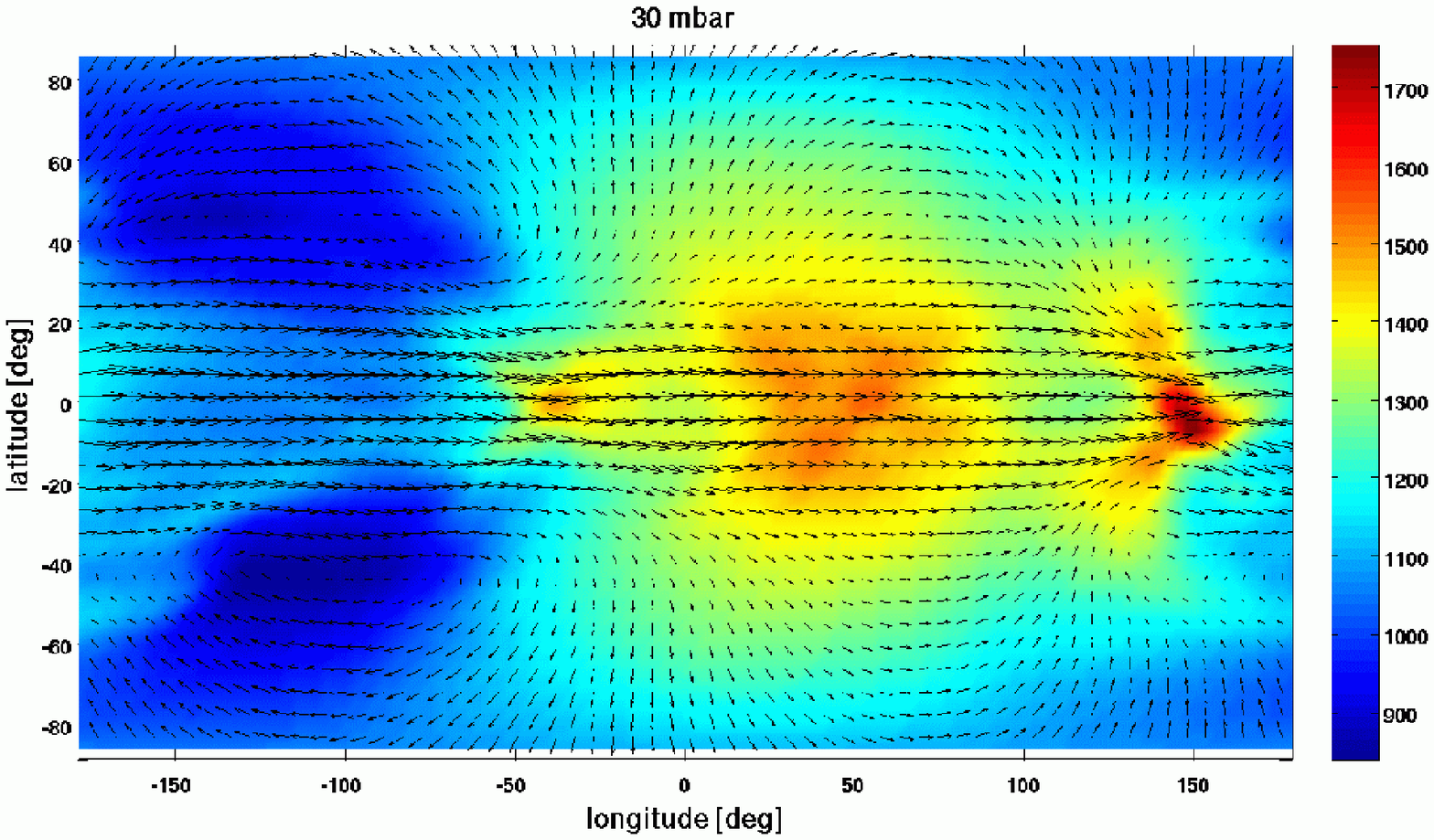}
\includegraphics[scale=0.43, angle=0]{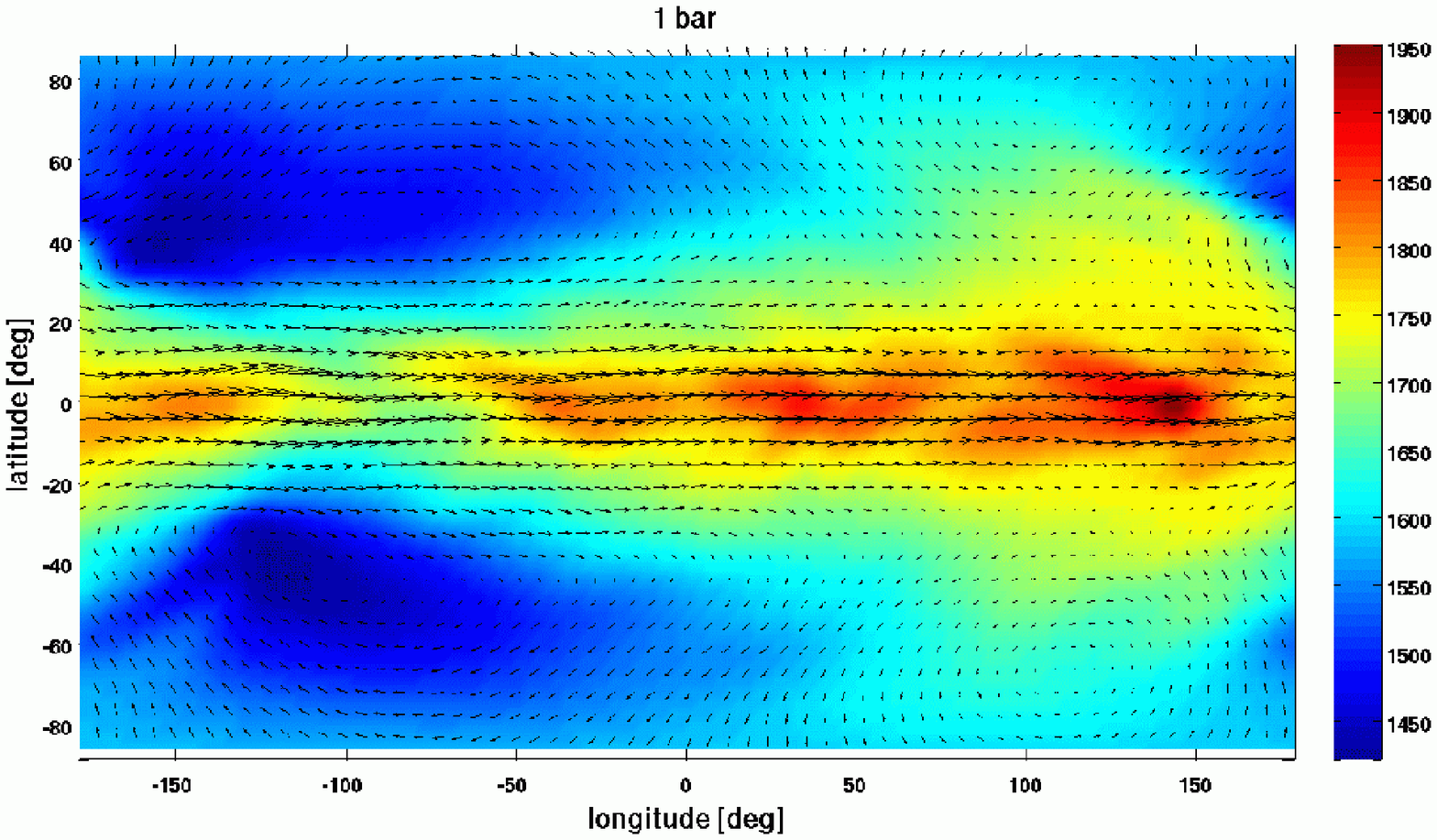}

\caption{Temperature (colorscale, in K) and winds (arrows) for nominal
HD 209458b simulation with solar abundances including TiO/VO.  Panels
show flow at 0.1 mbar ({\it top}), 1 mbar ({\it second panel}), 
30 mbar ({\it third panel}), and 1 bar ({\it bottom panel}).  Horizontal 
resolution is C32 (roughly equivalent to a resolution of $128\times64$
on a longitude/latitude grid) with 53 vertical layers.  
Substellar point is at longitude, latitude $(0^{\circ},0^{\circ})$.
Note the development of the dayside stratosphere.}
\label{hd209-solar}
\end{figure}

Our HD 209458b simulations develop vigorous circulations that, as expected,
include a hot dayside stratosphere.  This is illustrated in
Fig.~\ref{hd209-solar}, which shows the temperature 
and horizontal wind patterns at pressures of $1\,$bar, 30 mbar, 1 mbar, 
and 0.1 mbar ({\it bottom to top}) for our case with solar abundances.  
At deep levels (1 bar and 30 mbar), the circulation
qualitatively resembles that of HD 189733b, with an eastward equatorial
jet, westward high-latitude flows, and a hot region shifted east of
the substellar point (compare bottom two panels of Figs.~\ref{hd189-solar}
and \ref{hd209-solar}).  Nevertheless, even at these deep levels,
localized regions attain temperatures sufficient for TiO/VO to 
exist in the gas phase ($\sim1700$--$1900\,$K).

By pressures of 1 mbar, the picture changes drastically: a ``baby''
stratosphere with a radius of $50^{\circ}$ in longitude and latitude 
has developed, with temperatures reaching $2000\,$K (Fig.~\ref{hd209-solar}, 
{\it second panel}).  This stratosphere is approximately centered on the
substellar point, with an eastward offset of only $\sim10^{\circ}$ longitude.  
Its spatial confinement results from the fact that only air within 
$\sim50^{\circ}$ of the substellar point receives sufficient irradiation 
to achieve temperatures necessary for gas-phase TiO and VO to exist.
Air $>60^{\circ}$ from the substellar point, although still on the
dayside, has temperatures too low for gas-phase TiO/VO, and without
the benefit of the extra opacity supplied by these species, this
air remains substantially cooler ($<1400\,$K).  This nonlinear
feedback between temperature and opacity leads to a remarkably sharp 
temperature gradient, with temperatures dropping from 1800 to 1400 K 
as one moves from $50^{\circ}$ to $60^{\circ}$ angular distance from 
the substellar point.

At still lower pressures, an even greater fraction of the starlight
is available to cause heating, and the stratosphere becomes horizontally
larger until it covers most of the dayside at pressures $\le 0.1\,$mbar 
(Fig.~\ref{hd209-solar}, {\it top panel}). Although the equatorial winds 
at these low pressures involve a significant eastward flow at most
longitudes, the mid- and high-latitude winds involve a simpler motion that,
to zeroth order, moves air away from high-temperature
regions toward low-temperature regions.  At low pressure, the
temperature pattern is relatively symmetric about the substellar
and antistellar points, as suggested observationally for 
Ups And b \citep{harrington-etal-2006} and HD 179949 \citep{cowan-etal-2007}.
 
A contributing factor to the widening of the stratosphere with 
altitude is that stellar radiation is absorbed at lower pressure
near the limb than at the substellar point, which results directly
from the longer atmospheric path length to stellar irradiation at
the limb ($\mu\to0$) than at the substellar point ($\mu=1$).
Thus, near the limb the heating peaks at lower pressure, confining 
the stratosphere to low pressures. Near the substellar point, the 
heating peaks deeper, allowing the stratosphere to extend more
deeply in this localized region.

\begin{figure}
\vskip 10pt
\includegraphics[scale=0.38, angle=0]{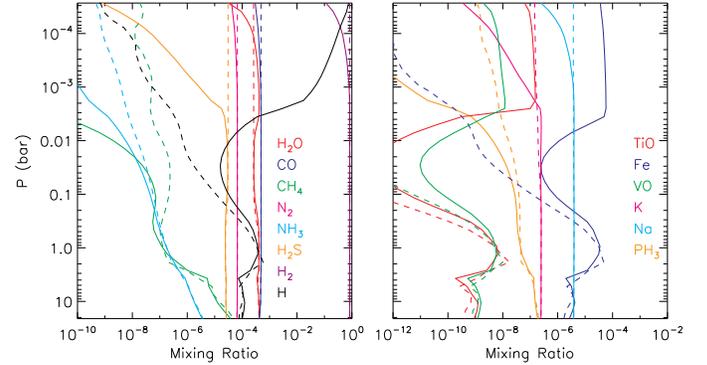}
\caption{Chemical-equilibrium abundances of several molecules versus 
pressure at the substellar point (solid) and antistellar point (dashed) 
from our three-dimensional HD 209458b case adopting solar abundances.  
}
\label{composition}
\end{figure}

Our simulated stratospheres result directly from the
existence of gaseous TiO and VO on the dayside in these simulations.  
Simulations of HD 209458b performed {\it without} gaseous TiO/VO lack 
stratospheres and develop temperature and wind patterns 
resembling a hotter version of our HD 189733b simulations.
Likewise, our previous HD 209458b simulations performed with
Newtonian heating/cooling \citep{showman-etal-2008a} developed
a weak dayside temperature inversion but nothing resembling
the extremely hot stratosphere shown in Fig.~\ref{hd209-solar}.

To illustrate the height-dependence of the composition (which
affects the opacity), Figure~\ref{composition} depicts the 
chemical-equilibrium abundances versus pressure at the substellar 
point (solid) and antistellar point (dashed) for our nominal HD 209458b 
case.  At low pressure, TiO is abundant on the dayside yet
depleted on the nightside.

Importantly, our simulated stratospheres do not extend
fully {\it to} the terminators (which are at longitudes $\pm90^{\circ}$ in 
Fig.~\ref{hd209-solar}).  Generally, the terminators themselves
have temperatures of $\sim1300\,$K or less.
The net heating there is simply too low
to allow temperatures above the TiO/VO condensation curves.  
Thus, while much of the dayside could have abundant TiO and VO, our
simulations suggest that these 
species will be largely absent (or at least depleted) on the limbs 
as seen during primary transit.  This will have important implications 
for interpreting transit spectra of HD 209458b \citep[e.g.][]{sing-etal-2008}
and other pM-class planets.  Interestingly, note that our simulated 
stratosphere approaches the teminator to the {\it east} of the substellar 
point more closely than it approaches the teminator to the {\it west} 
of the substellar point --- the result of thermal advection due to 
the eastward equatorial jet.  This result suggests that, during transit,
the planet's leading limb will be cooler and more depleted in TiO/VO than 
the trailing limb.

\begin{figure}
\vskip 10pt
\includegraphics[scale=0.55, angle=0]{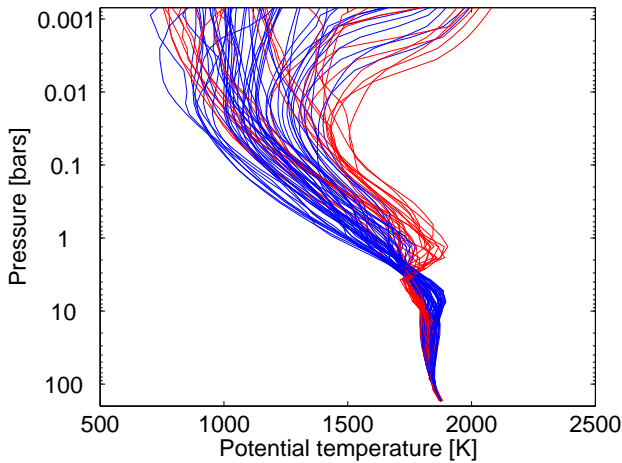}
\caption{A selection of temperature-pressure profiles for
our nominal, solar-abundance HD 209458b simulation including TiO
and VO opacity (same simulation as in Fig.~\ref{hd209-solar}).  
Red (blue) profiles are equatorward (poleward) of $30^{\circ}$ latitude.
Note the formation of the stratosphere at pressures less than $\sim30\,$mbar.}
\label{hd209-spaghetti}
\end{figure}

The diversity of temperature-pressure profiles in our solar model
is shown in Fig.~\ref{hd209-spaghetti}.  Near the bottom is the
quasi-isothermal region from $\sim3$--100 bars.  Above that, from
pressures of $\sim1\,$bar to 30 mbar, the temperature decreases
with altitude on both the dayside and the nightside.  At pressures
less than $\sim10\,$mbar, the dayside stratosphere is evident, 
with temperatures exceeding 2000 K.

\begin{figure}
\vskip 10pt
\includegraphics[scale=0.5, angle=0]{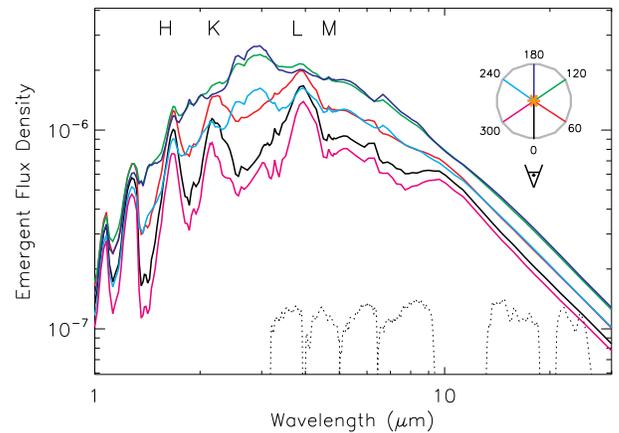}
\caption{Emergent flux density (${\rm ergs}\,\,\,{\rm s}^{-1}\, 
{\rm cm}^{-2}\,{\rm Hz}^{-1}$) from our nominal, solar-abundance
simulation of HD 209458b, including TiO and VO opacity, 
at six orbital phases.  {\it Black}, nightside,
as seen during transit; {\it red}, $60^{\circ}$ after transit;
{\it green}, $120^{\circ}$ after transit; {\it dark blue}, dayside,
as seen during secondary eclipse; {\it light blue}, $60^{\circ}$ after
secondary eclipse; and {\it magenta}, $120^{\circ}$ after secondary
eclipse. The key in the top right corner is color-coded with the spectra
to illustrate the sequence.  Thin dotted black lines at the bottom of
the figure show normalized {\it Spitzer} bandpasses and the letters at
the top show locations of the $H$, $K$, $L$, and $M$ bands.  This is
the same simulation as in Fig.~\ref{hd209-solar}.}
\label{hd209-spectra}
\end{figure}

Figure~\ref{hd209-spectra} shows spectra at six orbital phases.
Within $\sim60^{\circ}$ orbital phase of transit ({\it magenta, black, and red
curves}), molecular bands are seen in absorption because the 
temperature decreases with increasing altitude on the nightside.
Near secondary eclipse, however, when the dayside faces Earth,
the features flip into emission ({\it green, dark blue, and light
blue curves}), which results directly from the dayside temperature
inversion associated with the stratosphere.  This differs from
our HD 189733b simulations, where molecular features are seen
in absorption at all orbital phases, including the dayside 
(Fig.~\ref{hd189-spectra}).  Nevertheless, as the effective
temperature of this HD 209458b simulation makes clear (Fig.~\ref{Teff}),
the dayside radiation arrives primarily from the {\it bottom}
of the stratosphere where mean temperatures are a modest $\sim1500\,$K
(rather than from higher-altitude regions where temperatures
exceed $2000\,$K).

\begin{figure}
\vskip 10pt
\includegraphics[scale=0.5, angle=0]{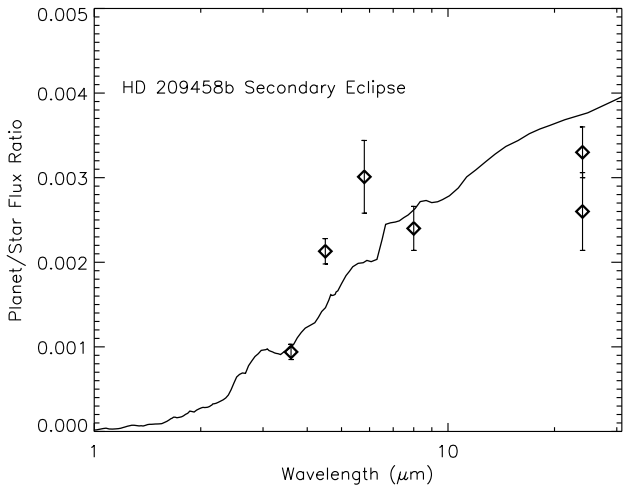}
\caption{Planet-to-star flux ratio vs. wavelength for our HD 209458b
simulation at the time immediately before/after secondary eclipse ({\it solid
curve}).  The simulation adopts solar abundances including TiO/VO opacity.
Points show measured secondary-eclipse depths \citep{deming-etal-2005a,
knutson-etal-2008a, knutson-etal-2009a}.
}
\label{hd209-planet-to-star}
\end{figure}

Figure~\ref{hd209-planet-to-star} compares our simulated HD 209458b
dayside planet/star flux ratio spectrum to the {\it Spitzer} secondary-eclipse 
photometry of \citet{knutson-etal-2008a, knutson-etal-2009a} and 
\citet{deming-etal-2005a}, including a tentative revision to the 
$24\,\mu$m eclipse depth by Deming (personal communication).  We
match the secondary-eclipse depths at 3.6 and $8\,\mu$m.  However,
despite the existence of a stratosphere in our simulations, our
solar-opacity case underpredicts the eclipse depths at 4.5 and 
$5.8\,\mu$m by nearly 50\% (roughly 3 and 2.3-$\sigma$, respectively, 
at these two bands).  We also miss the 24-$\mu$m eclipse 
depth of \citet{deming-etal-2005a} (the lower point in 
Fig.~\ref{hd209-planet-to-star}) by roughly 2.5-$\sigma$, 
although we fall within 2-$\sigma$ if the tentative revision to this 
point (upper point in Fig.~\ref{hd209-planet-to-star})
turns out to be more appropriate.

These discrepancies suggest that our simulated stratosphere
does not have the correct properties (e.g., temperature or
altitude range).  A comparison of our temperature profiles 
(Fig.~\ref{hd209-spaghetti}) to the radiative-equilibrium profiles
of \citet[][their Fig.~12]{fortney-etal-2008}, who match the 4.5 
and 5.8-$\mu$m points but overpredict the others, shows that
the hottest of our stratospheric profiles qualitatively resembles
theirs.  However, because our dayside includes a range of profiles
ranging from hot to cold, our {\it dayside average} profile is
colder than theirs.  This may help explain the fact that our predicted
dayside fluxes are systematically lower than those of 
\citet{fortney-etal-2008}.  Presumably, this difference results from
the vigorous dayside-to-nightside transport of thermal energy 
by the atmospheric circulation in our case.

When a spectrum exhibits greatly differing brightness temperatures 
at different wavelengths (as is true for HD 209458b), a common 
explanation is that the different wavelengths sense different pressure 
levels (because of the wavelength-dependent opacities).  In the presence 
of a vertically varying temperature this can produce a spectrum with 
wavelength-varying brightness temperatures.

In this context, a fundamental stumbling block to simultaneously explaining 
the five {\it Spitzer} secondary-eclipse depths is that the
range of pressures that contribute photons to the 3.6, 4.5, 5.6,
8, and 24-$\mu$m bands are all very similar --- at least for
the radiative-transfer model, chemical composition, 
and opacities adopted here.  
Contribution functions calculated for each of these {\it Spitzer} bands 
(see Fig.~\ref{contribution-fcns}, {\it right panel}) 
peak between 3 and 10 mbar, and they all have very broad tails
extending to $\sim0.1\,$mbar on the low-pressure side and
$\sim100\,$mbars on the high-pressure side.\footnote{These pressure 
ranges are specific to the solar model including TiO and VO.
Contribution functions calculated {\it without} TiO/VO for HD 189733b using
the same radiative-transfer code peak at substantially deeper pressures
of 30--$100\,$mbar.  See Fig.~\ref{contribution-fcns}.}   This overlap
in pressure means the brightness temperatures in all these bands tend to 
be similar.  Thus, it is difficult
to produce a high brightness temperature in some bands (such as at 4.5 
and $5.8\,\mu$m) while maintaining low brightness temperature in other
bands (such as at 3.6, 8, and $24\,\mu$m) --- as apparently required
by the data.  
This problem is not confined to the present study; it helps explain the
difficulty \citet{fortney-etal-2008} had in explaining all the
observations.  Even \citet{burrows-etal-2007b}, who had the flexibility
of several free parameters governing an assumed stratospheric absorber
and the magnitude and pressure range of the parameterized dayside heat sink,
were unable to match the 5.8/8-$\mu$m flux ratio.  Separating
the pressure ranges of the contribution functions would require differential
changes to the opacities at the wavelengths of {\it Spitzer} bandpasses.  
Potentially, alternate (i.e., disequilibrium) chemical compositions 
could sufficiently affect the opacities to resolve this conundrum;
investigating this possibility will require future work.  
Disequilibrium chemistry has been invoked to explain
infrared spectra of L dwarfs with similar effective 
temperatures \citep{leggett-etal-2007a}.

\begin{figure}
\vskip 10pt
\includegraphics[scale=0.5, angle=0]{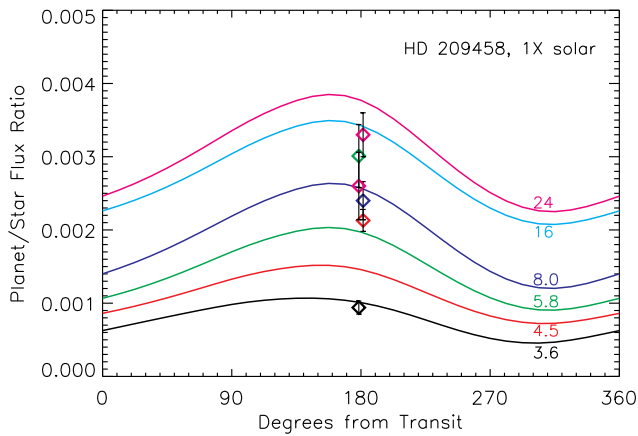}
\caption{
Light curves versus orbital phase calculated in {\it Spitzer} bandpasses
for our HD 209458b case with solar abundances.   
Within each panel, moving from bottom
to top, the light curves are for wavelengths $3.6\,\mu$m ({\it black}),
$4.5\,\mu$m ({\it red}), $5.8\,\mu$m ({\it green}), 
$8\,\mu$m ({\it dark blue}), $16\,\mu$m ({\it light blue}), 
and $24\,\mu$m ({\it magenta}), respectively.
Overplotted are the {\it Spitzer} secondary-eclipse measurements
from \citet{knutson-etal-2008a} in large diamonds, color-coded to
the light curves.}
\label{hd209-lightcurves}
\end{figure}

We now turn to infrared light curves.  Figure~\ref{hd209-lightcurves}
shows light curves in {\it Spitzer} bandpasses calculated for our
HD 209458b case with solar abundances.  As was the situation with HD 189733b,
our current HD 209458b light curves exhibit muted phase variations
relative to those obtained from our earlier simulations using
Newtonian heating/cooling \citep{showman-etal-2008a, fortney-etal-2006b}.
In the current simulations, the ratio of maximum-to-minimum 
planet/star flux ratio (at a given wavelength) is nearly 2 at all 
four IRAC bands
and $\sim1.6$--1.7 at the $16\,\mu$m IRS and $24\,\mu$m MIPS bands.
The {\it difference} between maximum 
and minimum values increases with wavelength and is 0.0014 at $8\,\mu$m. 
Phase offsets range from $\sim20^{\circ}$ at the longer wavelengths
to almost $40^{\circ}$ at $3.6\,\mu$m.  These values are significantly
nonzero but are smaller than their counterparts for our solar-abundance
HD 189733b simulations (Fig.~\ref{hd189-lightcurves}, {\it top panel}),
consistent with the general trend suggested by \citet{fortney-etal-2008}.

Our muted phase variations are consistent with the observational
constraint of \citet{cowan-etal-2007}, who found a 2-$\sigma$
upper limit of 0.0015 on the phase variation of HD 209458b at $8\,\mu$m.
Fig.~\ref{hd209-lightcurves} also re-iterates our agreement
with the 3.6 and $8\,\mu$m secondary-eclipse photometry
and our discrepancy at 4.5 and $5.8\,\mu$m.
Continuous light curves over half an orbit have recently been
obtained at 8 and $24\,\mu$m and are possible at 3.6 and $4.5\,\mu$m
with warm {\it Spitzer}, which would provide a test of our models
and insights on how to explain the discrepancies noted above.

The muted phase variations
in Fig.~\ref{hd209-lightcurves} are particularly intriguing given
that we {\it do} have a dayside stratosphere and (as a result) enormous
day-night temperature variations reaching $\sim1500\,$K
at low pressure. There are two possible reasons to reconcile why these
large day-night temperature variations do not translate into large
day-night flux variations.  First, much of the radiation that contributes
to the dayside fluxes emanates from altitudes near the bottom of the 
stratosphere, where day-night temperature differences are modest.  
This is exacerbated
by the fact that the stratosphere only covers part of the dayside.  Thus,
a substantial fraction of the dayside flux does not emanate from the
stratosphere but from deeper layers and surrounding regions 
that are cooler.  This lowers
the dayside flux compared to an idealized case where all the dayside
mid-infrared radiation comes from the stratosphere.
Second, the cooler temperatures on the nightside means that the 
mid-IR photospheres on the nightside are deeper in pressure --- by roughly 
an order-of-magnitude --- than those within the dayside stratosphere.  
Because temperatures increase with pressure
on the nightside, this elevates the nightside flux relative to
an idealized case where the photospheres remain at the same (low) pressure
everywhere. Both these effects act to mute the phase variations
in the mid-IR, at least in the {\it Spitzer} bandpasses.

Note, however, that the expected IR phase variations can
depend sensitively on wavelength.  As can be seen in 
Fig.~\ref{hd209-spectra}, for example, our HD 209458b
model predicts that the peak phase variations reach factors
of $\sim4$ in specific wavelength bands near 1.5, 1.8, and $2.8\,\mu$m,
with relatively smaller phase variations (factors of $\sim2$ or less) 
at intermediate wavelengths.    Since both large and small day-night 
phase variations can occur on a single object (depending on wavelength), 
caution is needed when attempting to link the amplitude of phase variation
to the efficiency of day-night heat transport --- especially if one
only possesses light curves at only one or two wavelengths.
Light curves obtained in numerous isolated 
bands across this region (or low-to-moderate resolution spectra at
different orbital phases) by a future space mission could shed 
considerable insight into the atmospheric composition and 3D temperature 
structure.

\subsection{Discussion}

Our results point toward possible
refinements of the scenario outlined in \citet{fortney-etal-2008}.
They anticipated large day-night flux
differences for pM-class planets, including HD 209458b.  This contrasts
with the modest phase variations we find here. 
However, the fact that we underpredict 
the dayside 4.5 and 5.8-$\mu$m fluxes for HD 209458b 
suggests that {\it in reality} 
(as opposed to in our simulations) the day-night flux variations will be
large at these two wavelengths.  If similar flux ratios are found for 
other planets with day-side temperature inversions \citep[XO-1b may point to 
this; see][]{machalek-etal-2008}, a revision to the \citet{fortney-etal-2008} 
scenario may be in order, wherein HD 209458b and other transitional pM-class
planets will have large phase variations at 4.5 and $5.8\,\mu$m but
modest phase variations at the other Spitzer bandpasses. $K$
band, sensing even deeper than $\sim100\,$ mbar, should also exhibit
only modest phase variations.  This scenario reconciles
the secondary-eclipse spectrum of HD 209458b with the observational
upper limit on the phase variation at $8\,\mu$m \citep{cowan-etal-2007}.

But what of hotter planets? \citet{fortney-etal-2008} positioned HD 209458b
near the pL/pM boundary and acknowledged that this transition
could be indistinct because of the extended temperature range over
which TiO and VO condense \citep{lodders-2002}.
Planets more strongly illuminated than HD 209458b 
should have hotter stratospheres
that cover more of the dayside, and it remains possible that
such ``very hot Jupiters'' could indeed have large day-night phase 
variations in most {\it Spitzer} channels as predicted by
\citet{fortney-etal-2008}. The strong observational evidence that 
ups And b \citep{harrington-etal-2006} and HD 179949b \citep{cowan-etal-2007} 
show large day-night flux contrast at 24 and $8\,\mu$m, respectively, 
may indicate such a change in temperature structure at the higher 
incident fluxes ($\sim35\%$ higher than HD209458b) that these planets 
receive.   Although neither of these planets eclipse, 
\citet{barnes-etal-2008} have suggested that HD 179949b has a dayside 
temperature inversion.
It will be interesting to perform 3D simulations of these and
hotter planets, particularly ones that go through secondary eclipse.
Obtaining light curves for very hot eclipsing systems, such
as TrES-4 and HAT-P-7b, should be important goals for warm {\it Spitzer}
or JWST.

 For all of these systems, whether large day-night flux contrasts in the 
IR translate into large day-night effective temperature contrasts 
can only be unambiguously answered by light curves across many wavelengths, 
preferably from 2--$5\,\mu$m, around the peak in planetary flux 
\citep[e.g.][]{barman-2008}.  This makes 3.6 and $4.5\,\mu$m 
light curves from the {\it Spitzer} warm mission, and continued searches
for planet flux in $K$ band, particularly important.

Finally, we conclude this section with some discussion of the likelihood 
that gaseous TiO and VO can actually exist on the dayside.  As discussed 
in \citet{fortney-etal-2008}, we calculate opacity
assuming {\it local} chemical equilibrium at the given $p$ and $T$.
Assuming TiO and VO are included in the database in the first place, 
this means that gaseous TiO/VO opacity are included if the local
temperatures are hot enough --- ignoring the possibility of any
``cold trap'' effect that could globally deplete the abundance of
these species.  One possible cold trap, commonly discussed in the
context of 1D models, is at pressures of tens to hundreds of bars 
where the temperature at the bottom of the near-isothermal radiative
zone is expected to be cooler than the TiO/VO condensation temperatures for
multi-Gyr-old hot Jupiters farther than $\sim0.04\,$AU from their
stars \citep{hubeny-etal-2003,
fortney-etal-2008}. Nevertheless, the existence of stratospheres 
on some planets argues that this does not occur on HD 209458b and 
other pM-class planets. 

However, another possible cold trap --- not considered by previous 1D
models --- is the presence of the large day-night temperature difference
in the observable atmosphere.  As a parcel of hot dayside air flows
onto the nightside, its temperature plummets and gaseous TiO and VO
condense.   If these Ti- and V-bearing particles settle out before
the air parcel returns to the dayside, then the atmosphere could become
depleted in TiO and VO even if no cold trap exists at deeper levels
of tens to hundreds of bars.  On the other hand, if the particles
are small, their settling speeds will be slow and they will remain 
in the air parcel when it returns to the dayside days later.  In
this case, the increased temperatures will allow these particles to
sublimate, resupplying gaseous TiO and VO to the atmosphere.  Thus,
our simulations with TiO/VO are only self-consistent if the 
TiO/VO particles cannot settle out on the nightside.  This effectively
means that the particles must remain small.

We can quantify the maximum particle sizes as follows.  At 
1 mbar, characteristic vertical velocities in our HD 209458b simulation
are $\sim30\,{\rm m}\,{\rm s}^{-1}$.  For particles to remain
suspended, the settling velocities must be less than these values.
Given the Stokes flow speed modified for gas-kinetic effects
\citep[see e.g.][Eq.~B1]{ackerman-marley-2001}, and air viscosities
relevant to hydrogen at $\sim1500\,$K, this implies particles radii
less than $\sim30\,\mu$m.  At lower pressure, the gas-kinetic effects
become stronger, leading to greater settling velocities; at 
0.1 mbar, for example, the particles must be less than 
$\sim7$--$10\,\mu$m in radius to remain suspended.   (Calculation
of Reynolds numbers for the falling particles shows that they
are less than one, implying that the Stokes relation is valid
and that turbulent modifications to the fall velocity
need not be considered.)  It seems plausible that the actual particles
sizes are smaller than these values, but detailed microphysical 
calculations will be needed to explore this further.  A 
complicating factor is that not only TiO/VO but also silicates
will condense on the nightside at low pressure.

These estimates are crudely
consistent with subsequent calculations performed by \citet{spiegel-etal-2009b},
who parameterized the dynamics in 1D using eddy diffusion and estimated 
the eddy-diffusion coefficients needed to keep TiO particles
of various sizes lofted.  If one translates our vertical velocities
into eddy diffusivities by multiplying the vertical velocities by
the pressure scale height, then the implied eddy diffusivities at 1 mbar
are $\sim10^{11}\rm cm^2 \,sec^{-1}$.  In their 
HD 209458b model, at 1 mbar, these eddy diffusivities are sufficient to loft particles 
smaller than several $\mu$m.  Thus, their calculations 
likewise suggest that TiO stratospheres may be viable if the TiO particle 
sizes are less than several $\mu$m.

\section{Conclusions}

We presented global, three-dimensional numerical simulations
of the atmospheric circulation of HD 189733b and HD 209458b that 
couple the dynamics to a realistic representation of cloud-free
non-gray radiative transfer.  This new model, which we dub SPARC/MITgcm,
is the first 3D dynamical model 
for any giant planet --- including those in our Solar System --- 
to incorporate nongray radiative transfer.
Our model adopts the MITgcm for the dynamics
and an optimized version of the radiative model of McKay, Marley, Fortney,
and collaborators for the radiative transfer.  Opacities are 
calculated assuming solar composition (or some multiple thereof)
with equilibrium chemistry that accounts for rainout.  Like earlier
work with simplified forcing \citep{showman-etal-2008a,
cooper-showman-2005, showman-guillot-2002}, our simulations develop 
a broad eastward equatorial jet, mean westward flow at high latitudes, 
and substantial flow over the poles at low pressure.  The jet structure 
depends significantly on longitude at pressures $<100\,$mbar.

For HD 189733b, our simulations that exclude TiO and VO opacity
explain the broad features of the observed 8 and 24-$\mu$m
light curves \citep{knutson-etal-2007b, knutson-etal-2009a}, 
including the modest day-night flux variation
and the fact that the planet/star flux ratio peaks before the 
secondary eclipse.  In our simulations, the offset results from the
eastward displacement of the hot regions from the substellar point.
On the other hand, we do not fit the flux minimum seen after
transit in the 8-$\mu$m light curve \citep{knutson-etal-2007b}.
Our simulations also provide reasonable matches to the 
{\it Spitzer} secondary-eclipse depths at 4.5, 5.8, 8, 16, 
and $24\,\mu$m \citep{charbonneau-etal-2008, deming-etal-2006}
and the groundbased upper limit at $2.2\,\mu$m from
\citet{barnes-etal-2007}.  The temporal variability in these
simulations is modest --- of order 1\% --- and is fully consistent
with the upper limit on temporal variability from \citet{agol-etal-2008}.

The primary HD 189733b observation where we fare poorly is the
$3.6\,\mu$m secondary-eclipse depth from \citet{charbonneau-etal-2008},
which we underpredict by about a factor of two.  Because the
$3.6\,\mu$m channel is expected to sense down to $\sim0.1$--1 bar
on this planet (Fig.~\ref{contribution-fcns}, {\it left panel}), 
this suggests that 
our simulation is too cold in this region of the atmosphere and/or
has the incorrect opacity in this wavelength range.  Previous
1D models of HD 189733b have suffered a similar problem at this wavelength
\citep{barman-2008, knutson-etal-2009a}, as have 1D models of
brown dwarfs with similar effective temperatures \citep{geballe-etal-2009}.
A full-orbit light curve of HD 189733b at $3.6\,\mu$m, possible with 
warm {\it Spitzer}, would provide crucial insights to help resolve 
this problem.

For HD 209458b, we include gaseous TiO and VO opacity to see
whether it allows us to explain the inference of a stratosphere from 
{\it Spitzer} photometry \citep{knutson-etal-2008a, burrows-etal-2007b, 
fortney-etal-2008}.  As expected, these simulations
develop a hot ($>2000\,$K) dayside stratosphere whose horizontal
dimensions are small at depth but widen with altitude until
the stratosphere covers most of the dayside at pressures $<0.1\,$mbar.
Interestingly, both branches of the bifurcation discussed by 
\citet{hubeny-etal-2003} for different planets occur here on a 
single planet's dayside.  Using the terminology of \citet{fortney-etal-2008},
the substellar region is a pM-class planet
but the terminators and nightside are a pL-class planet.  It
is thus perhaps more proper to talk about pM-class {\it daysides}
rather than pM-class {\it planets}.

But despite the stratosphere in our simulations of HD 209458b, 
we do not reproduce current {\it Spitzer} photometry of this planet,
which includes particularly high ($\sim1700$--$1900\,$K) brightness
temperatures in the 4.5 and 5.8-$\mu$m channels.  This could 
mean that our stratosphere has the incorrect properties (e.g.,
temperature range, altitude range, and vertical thermal gradient).
However, a fundamental difficulty in explaining the diverse 
brightness dayside temperatures (ranging from $\sim1500$--$1900\,$K
in {\it Spitzer} bandpasses) is that the range of pressures from 
which emergent infrared photons originate are very similar for all 
{\it Spitzer} bandpasses, at least as calculated by our radiative
model with equilibrium chemistry.   This means that {\it regardless} 
of the planet's temperature 
profile, the brightness temperatures in all these bands should be similar. 
Breaking this degeneracy would require changing the opacities so that
the opacities in different {\it Spitzer} bandpasses differ significantly.
Dropping the equilibrium chemistry assumption would make this possible;
this provides a clue that disequilibrium chemistry may be important.
Disequilibrium chemistry appears to influence infrared spectra of brown
dwarfs \citep[e.g.][]{leggett-etal-2007a, geballe-etal-2009}, lending 
weight to this possibility.

Our light curves of HD 209458b in {\it Spitzer} bandpasses exhibit
modest day-night variation, and we successfully explain the upper 
limit on the day-night flux contrast from \citet{cowan-etal-2007} at 
$8\,\mu$m.  Our ability to meet this constraint results from the 
fact that much of the dayside 8-$\mu$m radiation emanates from
altitudes near the base of the stratosphere, where temperatures are not too
hot, while nightside 8-$\mu$m radiation emanates from substantially
deeper pressures where temperatures are warmer than they are aloft.  
This dual effect leads to modest day-night flux variations despite 
large day-night temperature variations (on isobars) at low pressures.  
Assuming the high inferred 4.5 and 5.8-$\mu$m brightness temperatures indeed
result from a stratosphere, we suggest that the real planet should
exhibit large phase variations at 4.5 and $5.8\,\mu$m yet more modest
phase variations at other {\it Spitzer} bandpasses and $K$ band.

The task of developing exoplanet
GCMs that couple dynamics and radiative transfer has now been espoused 
by many authors, and the work presented here demonstrates that
this approach indeed holds significant promise for 
explaining atmospheric observations of hot Jupiters.  Our 3D 
SPARC/MITgcm simulations, 
especially of HD 189733b, show encouraging resemblance to observations of the 
real planet.  While some discrepancies remain, and a wider range of
parameters need be explored, these simulations support the idea
that detailed model/data comparisons can eventually allow robust inferences
about the circulation regime of hot Jupiters to be inferred.  
Given the huge range in properties of currently known transiting
planets, with future observations sure to unveil additional surprises,
this helps energize the exciting prospect that planetary meteorology 
can successfully be extended beyond the confines of our 
Solar System.


\acknowledgements
This research was supported by NASA Origins grant NNX08AF27G
and Planetary Atmospheres grants NNX07AF35G and NNG06GF28G to APS. 



\end{document}